\documentclass[%
superscriptaddress,
 amsmath,amssymb,
 aps,
]{revtex4-2}

\usepackage{hyperref}
\usepackage{graphicx}
\usepackage{dcolumn}
\usepackage{bm}
\usepackage{amsmath}
\usepackage{color}

\usepackage{lineno}
\newcommand*\patchAmsMathEnvironmentForLineno[1]{
  \expandafter\let\csname old#1\expandafter\endcsname\csname #1\endcsname
  \expandafter\let\csname oldend#1\expandafter\endcsname\csname end#1\endcsname
  \renewenvironment{#1}
     {\linenomath\csname old#1\endcsname}
     {\csname oldend#1\endcsname\endlinenomath}}
\newcommand*\patchBothAmsMathEnvironmentsForLineno[1]{
  \patchAmsMathEnvironmentForLineno{#1}
  \patchAmsMathEnvironmentForLineno{#1*}}
\AtBeginDocument{
\patchBothAmsMathEnvironmentsForLineno{equation}
\patchBothAmsMathEnvironmentsForLineno{align}
\patchBothAmsMathEnvironmentsForLineno{flalign}
\patchBothAmsMathEnvironmentsForLineno{alignat}
\patchBothAmsMathEnvironmentsForLineno{gather}
\patchBothAmsMathEnvironmentsForLineno{multline}
}

\begin{document}

\preprint{APS/123-QED}

\title{Intrinsic regularization effect in Bayesian nonlinear regression scaled by observed data}
\author{Satoru Tokuda}
\email{s.tokuda.a96@m.kyushu-u.ac.jp}
\affiliation{Research Institute for Information Technology, Kyushu University, Kasuga, Fukuoka 816-8580, Japan}
\affiliation{Mathematics for Advanced Materials -- Open Innovation Laboratory, AIST, c/o Advanced Institute for Materials Research, Tohoku University, Sendai, Miyagi, 980-8577, Japan}
\affiliation{Department of Complexity Science and Engineering, The University of Tokyo, Kashiwa, Chiba 277-8561, Japan}
\author{Kenji Nagata}
\affiliation{Department of Complexity Science and Engineering, The University of Tokyo, Kashiwa, Chiba 277-8561, Japan}
\author{Masato Okada}%
\affiliation{Department of Complexity Science and Engineering, The University of Tokyo, Kashiwa, Chiba 277-8561, Japan}
\affiliation{Center for Materials Research by Information Integration, National Institute for Materials Science, Tsukuba, Ibaraki 305-0047, Japan}
\date{\today}

\begin{abstract}
Occam's razor is a guiding principle that models should be simple enough to describe observed data. While Bayesian model selection (BMS) embodies it by the intrinsic regularization effect (IRE), how observed data scale the IRE has not been fully understood. In the nonlinear regression with conditionally independent observations, we show that the IRE is scaled by observations' fineness, defined by the amount and quality of observed data. We introduce an observable that quantifies the IRE, referred to as the Bayes specific heat, inspired by the correspondence between statistical inference and statistical physics. We derive its scaling relation to observations' fineness. We demonstrate that the optimal model chosen by the BMS changes at critical values of observations' fineness, accompanying the IRE's variation. The changes are from choosing a coarse-grained model to a fine-grained one as observations' fineness increases. Our findings expand an understanding of BMS's typicality when observed data are insufficient.
\end{abstract}

\pacs{Valid PACS appear here}
\maketitle

\section{Introduction}
To describe observed data by mathematical model is an important stage to understand the physics of the system behind observed data.
It is desired to build a model as simple as possible for caricaturing the essential physics \cite{goldenfeld1992lectures, goldenfeld1999simple, batterman2002asymptotics, batterman2014minimal, oono2012nonlinear}.
A simple model is an equation or a function with fewer parameters, which sometimes corresponds to an effective theory: a special case, an approximation, or a coarsening of the more comprehensive theory.
Some attempts relate the model simplicity to a kind of {\it emergence} as if microscopic details are almost negligible in macroscopic phenomena from the viewpoint of information theory \cite{hoel2013quantifying, machta2013parameter, mattingly2018maximizing, gordon2021relevance}. They are also related to what model is simple enough for describing observed data well.
In the context of statistics, some criteria for model selection, as highlighted by the Akaike and Bayesian information criteria (AIC and BIC), support selecting a simpler model as justifications of Occam's razor \cite{akaike1974new, schwarz1978estimating}.

Bayesian model selection (BMS) is utilized to choose a model simple enough to describe the essential physics behind observed data \cite{trotta2008bayes, mann2011bayesian, mark2018bayesian, vazquez2021bayesian, tokuda2021unveiling, tokuda2021bayesian}.
Suppose that the ground truth that generates the observed data is included in the list of candidate models.
There are several motivations for employing the BMS. 
One is that the BMS is consistent \cite{wasserman2000bayesian, berger2001objective}. If the candidate models are disjoint, the model chosen by the BMS converges almost surely toward the ground truth as observed data increase under mild conditions.
Another is that the BMS automatically incorporates Occam's razor \cite{wasserman2000bayesian, berger2001objective}.
If more than one candidate model can exactly express the ground truth, the model chosen by the BMS converges almost surely toward the simplest one with the fewest parameters. We should mention that our primary interest in the BMS is choosing the simplest model that describes observed data well rather than verifying one's model (hypothesis) based on the data. The simplest model for given data is sometimes consistent with the ground truth, but not necessarily. We want to clarify when and how the simplest model is the ground truth or other candidates.

Occam's razor in the BMS is embodied by the intrinsic regularization effect (IRE) that prefers simpler models to complex ones as succeeding to the BIC \cite{schwarz1978estimating, mackay1992bayesian, balasubramanian1997statistical}.
In the case of enough observed data, the singular learning theory has revealed that the IRE is explicitly quantified by a birational invariant, called the real log canonical threshold (RLCT) \cite{watanabe2001algebraic, watanabe2009algebraic, watanabe2018mathematical}.
However, how observed data scale the IRE is still a challenging question. This question includes three issues.
First, the RLCT does not quantify the dependence on the amount and quality of observed data.
We need the IRE's scaling function that converges toward the RLCT in the limit where they are sufficient.
Second, the RLCT is not an observable quantity calculated from observed data but a quantity theoretically derived from the ground truth, which is practically unknown. 
While some observable quantities that converge toward the RLCT have been introduced \cite{watanabe2013widely, tokuda2014numerical}, they do not explicitly represent the dependence on both the amount and quality of observed data. We also need an observable counterpart of the scaling function.
Third, how the IRE affects the BMS in the case of insufficient observed data has yet to be fully understood. While the BMS is consistent, the model chosen by the BMS is not necessarily the ground truth if observed data are insufficient. We also need to explain the typicality of the BMS for insufficient data by using the scaling function.

Here we address these issues in the context of the nonlinear regression with conditionally independent observations, which is a prototypical setup in mathematical modeling.
To quantify the IRE, we introduce an observable quantity, referred to as the Bayes specific heat, inspired by the mathematical equivalence between statistical inference and statistical physics \cite{jaynes1957information, balasubramanian1997statistical, jaynes2003probability, zdeborova2016statistical}.
We derive its finite-size scaling relation to observations' fineness, defined by the amount and quality of observed data. We show the correspondence of the scaling function to the RLCT. 
We demonstrate that the model chosen by the BMS changes at critical values of observations' fineness, affected by variation in the scaling function. We also find that the changes are from choosing a coarse-grained model to the fine-grained ground truth as observations' fineness increases. Our findings correspond to a typical behavior of the BMS with a variation of observed data from insufficient to sufficient, expanding an understanding of the BMS's consistency.

\section{Statistical ensemble of nonlinear regression}
\label{sec:Sec.2}
We start by defining a {\it statistical ensemble} of nonlinear regression. Let us consider observing conditionally independent random variables $D^n:=\{x_i, Y_i\}_{i=1}^n$ for $x_i = x_{i-1} + L/(n-1)$, where $x_1$, $x_n$, and $L := x_n-x_1> 0$ are fixed.
We assume that $Y_i \sim \mathcal{N}(f(x_i;w), \beta^{-1})$, where $f$, $w$, and $\beta^{-1} > 0$ are respectively regression model, its parameter set, and variance of observation noise as an embodiment of the quality of observed data.
Here we define observations' fineness $\kappa :=n \beta$, which is a key quantity of scaling relations as described later.
Since $w$ and $f$ are unknown and should be estimated from $D^n$, we treat them as random elements subject to the posterior probability distribution
\begin{align}
p(w, f \mid  D^n, \beta, \mu) \propto \exp \left( -\frac{\kappa}{2} E_n(w;f) - \kappa \mu F_n(\beta,f) \right) p(w \mid f) p(f),
\label{eq:ensemble}
\end{align}
where
\begin{align}
E_n(w;f) := \frac{1}{n} \sum_{i=1}^n (Y_i - f(x_i;w))^2
\label{eq:def_E}
\end{align}
is the mean square error, 
\begin{align}
F_n(\beta, f) := - \frac{1}{\kappa} \log \int \exp \left( -\frac{\kappa}{2} E_n(w;f) \right) p(w \mid f) dw
\label{eq:def_F}
\end{align}
is the Bayes free energy, $\mu \geq 0$ is an auxiliary variable, $p(w \mid f)$ is an arbitrary prior probability distribution of $w \in W \subset \mathbb{R}^d$, and $p(f)$ is that of $f \in \{ f^K \}$. 
If $\mu=1$, Eq. \eqref{eq:ensemble} is derived from Bayes' theorem. Throughout this study, we consider the case $\mu \to \infty$, which corresponds to the empirical Bayes approach, i.e., $f$ is chosen by minimizing $F_n(\beta, f)$ at a given $\beta$ as the BMS \cite{efron1973stein, akaike1998likelihood, mackay1992bayesian}. Note that the BIC and its generalized version are derived by approximating $F_n(\beta, f)$ \cite{schwarz1978estimating, watanabe2013widely}. While $\beta$ is given in Eq. \eqref{eq:ensemble}, if $\beta$ is unknown, it can also be estimated by minimizing $\kappa F_n(\beta, f) - n (\log \beta -\log 2 \pi)/2$ \cite{tokuda2017simultaneous}. The overall setup is the basics of our analyses in the following sections to clarify the BMS transitions from choosing a coarse-grained model to the fine-grained ground truth as $\kappa$ increases.

\section{Bayes specific heat}
\label{sec:Sec.3}
We introduce a key quantity that characterizes the {\it macrostates} in the statistical ensemble of nonlinear regression, which is defined as the fluctuation of $E_n(w;f)$ at a certain pair of $\beta$ and $f$ with a given $D^n$: 
\begin{align}
C_n(\beta, f) &= \left( \frac{\kappa}{2} \right)^2 \left( \langle E_n(w; f)^2 \rangle - \langle E_n(w; f) \rangle^2 \right),
\label{eq:SH}
\end{align}
where $\langle \cdots \rangle := \int (\cdots) p(w \mid \beta, f, D^n) dw$ denotes the average over all {\it microstates} of $w$ subject to $p(w \mid \beta, f, D^n) \propto \exp ( -\kappa E_n(w;f) /2 ) p(w \mid f)$. Note that Eq. \eqref{eq:SH} is derived from a more general definition of such a quantity in statistical inference, called the Bayes specific heat (see Appendix \ref{sec:App.A}). 
We should mention that $C_n(\beta, f)$ is an observable that quantifies the IRE as explained in the next section. This quantity plays an important role in explaining the mechanism of the BMS transitions in terms of the IRE.

\section{Finite-size scaling relations}
We hereafter consider the case that $Y_i \sim \mathcal{N}(f_0(x_i;w_0), \beta_0^{-1})$, where $f=f_0$, $w=w_0$ and $\beta=\beta_0$ are the ground truths such that $p(w_0 \mid f_0) >0$ is satisfied. 
The BMS depends on the set $D^n$ of $n$ random variables since it is carried out by minimizing $F_n(\beta, f)$, which is a function of $D^n$. Besides, $F_n(\beta, f)$ depends on both $D^n$ and $\beta$ rather than on $\kappa$ (see Eqs. \eqref{eq:def_E} and \eqref{eq:def_F}). To clarify the typicality of BMS transitions, not depending on the realization of $D^n$ but on $\kappa$, we take the limit where $\kappa$ and $L$ are fixed, whereas $n \to \infty$. We derive the finite-size scaling relation
\begin{align}
F_n(\beta, f) &= F(\kappa, f; w_0, f_0) + \frac{R_n}{2 n \beta_0}
\label{eq:SFF}
\end{align}
with the scaling function
\begin{align}
F(\kappa, f; w_0, f_0) :&= - \frac{1}{\kappa} \log \int \exp \left(- \frac{\kappa}{2} E(w;f, w_0, f_0) \right) p(w \mid f) dw
\label{eq:SFF2}
\end{align}
and the {\it energy} function
\begin{align}
E(w;f, w_0, f_0) := \frac{1}{L} \int_{x_1}^{x_n} \left( f_0(x ;w_0) - f(x ;w) \right)^2 dx,
\end{align}
where $R_n$ is a random variable subject to the chi-square distribution with $n$ degree of freedom (see Appendix \ref{sec:App.B}). Note that $F_n(\beta, f)$ is self-averaging, i.e., $F_n(\beta, f) = [F_n(\beta, f)]$ holds as $n \to \infty$, where $[\cdots] :=  \int (\cdots) \prod_{i=1}^n p(y_i \mid x_i, w_0, \beta_0, f_0) d y_i$ denotes the average over all realizations of $D^n$, which are subject to $p(y_i \mid x_i, w_0, \beta_0, f_0) := \mathcal{N}(f_0(x_i;w_0), \beta_0^{-1})$. 
Since $R_n / (2 n \beta_0)$ is independent of $f$, $f$ minimizing $F(\kappa, f; w_0, f_0)$ is equal to $f$ minimizing $F_n(\beta, f)$ at $\kappa$ consisting of a certain pair of $\beta$ and sufficiently large $n$. Under the condition $\beta=\beta_0$, corresponding to the Nishimori line \cite{nishimori1980exact, iba1999nishimori}, $F(\kappa, f; w_0, f_0)$ enables us to assess the typical behavior of statistical ensemble at a certain $\kappa=n\beta_0$, which is independent of the realization of $D^n$. Based on this typicality, we demonstrate that $f$ minimizing $F(\kappa, f; w_0, f_0)$ at any $\kappa$ is not necessarily $f_0$ in the next section. The demonstration clarifies the typicality of BMS transitions, independent of the realizations of $D^n$, from choosing a coarse-grained model to the fine-grained ground truth as $\kappa$ increases.

We are interested in explaining the mechanism of BMS transitions in terms of the IRE. To explain the mechanism formally, in the above limit where $\kappa$ and $L$ are fixed, whereas $n \to \infty$, we also derive the finite-size scaling relation
\begin{align}
C_n(\beta, f) = \Lambda(\kappa, f; w_0, f_0) + O_p \left( \max \left( \frac{1}{\log \kappa}, \frac{\beta}{\beta_0}, \frac{1}{n \sqrt{\beta_0}} \right) \right)
\label{eq:FSS2}
\end{align}
with the scaling function
\begin{align}
\Lambda(\kappa, f; w_0, f_0) :&= \left( \frac{\kappa}{2} \right)^2 \left( \langle E(w;f, w_0, f_0)^2 \rangle - \langle E(w;f, w_0, f_0) \rangle^2 \right),
\label{eq:RLCT}
\end{align}
where $\langle \cdots \rangle$ converges to the average over $p(w \mid \beta, f, D^n) \propto \exp ( -\kappa E(w;f, w_0, f_0) /2 ) p(w \mid f)$ in the limit we consider (see Appendix \ref{sec:App.C}).
From Eq. \eqref{eq:FSS2}, we find that $C_n(\beta, f)$ is an observable counterpart of $\Lambda(\kappa, f; w_0, f_0)$ in the limit we consider.
Note that $C_n(\beta, f)$ is conditionally self-averaging, i.e., $C_n(\beta,f) = [C_n(\beta,f)]$ holds as $n \to \infty$ if $\beta=O(\beta_0 / \log n)$.
Thus, $C_n(\beta,f)$ is not necessarily self-averaging under the condition $\beta=\beta_0$, i.e., $C_n(\beta_0, f) = \Lambda(n \beta_0, f; w_0, f_0) + O_p \left( 1 \right)$ holds.
We assess the finite-size effect on some particular examples to validate that the fluctuation term $O_p \left( \max \left( (\log \kappa)^{-1}, \beta/\beta_0, (n \sqrt{\beta_0})^{-1} \right) \right)$ is negligible at any $\beta$ in some situations (see Appendix \ref{sec:App.D}).

We also find that $\Lambda(\kappa, f; w_0, f_0)$ is a quantification of the IRE in the limit we consider from the consideration below. There is a junction between Eq. \eqref{eq:RLCT} and the singular learning theory \cite{watanabe2001algebraic, watanabe2009algebraic, watanabe2010equations, watanabe2010asymptotic, watanabe2013widely, watanabe2018mathematical}.
While the setup is based on independent and identically distributed observations as $n \to \infty$ in the related study \cite{watanabe2010limit}, our setup is based on conditionally independent observations in the limit where $\kappa$ and $L$ are fixed, whereas $n \to \infty$. Our setup is a natural extension,  where the limits of $F(\kappa, f; w_0, f_0)$ and $\Lambda(\kappa, f; w_0, f_0)$ as $\kappa \to \infty$ converge to the related study's result. 
Any direct counterparts of $C_n(\beta, f)$ and $\Lambda(\kappa, f; w_0, f_0)$ have not been defined in the singular learning theory, while Eq. \eqref{eq:FSS2} is related to Watanabe's corollary \cite{watanabe2013widely} (see Appendices \ref{sec:App.A} and  \ref{sec:App.C}).
Thanks to this relation, the limit of $\Lambda(\kappa, f; w_0, f_0)$ as $\kappa \to \infty$ is regarded as the RLCT, which characterizes the IRE as the coefficient of a leading term in the asymptotic expansion of $F_n(\beta, f)$ as $n \to \infty$ \cite{watanabe2001algebraic, watanabe2009algebraic, watanabe2018mathematical} (see Eq. \eqref{eq:leading}).
Namely, $\Lambda(\kappa, f; w_0, f_0)$ quantifies the IRE not only in the limit $n \to \infty$ but also in the limit where $\kappa$ and $L$ are fixed, whereas $n \to \infty$.
The limit $\Lambda(\kappa, f; w_0, f_0) \to {\rm dim}(w)/2$ also holds as $\kappa \to \infty$ if $p(w \mid D^n, \beta, f)$ is regular, where the BIC is justified \cite{schwarz1978estimating, watanabe2013widely}.
Based on these correspondences, we demonstrate how the IRE is scaled by $\kappa$ and how it affects the BMS transitions in the next section. 

\section{Intrinsic regularization effect scaled by observed data}
We demonstrate the BMS transitions from choosing a coarse-grained model to the fine-grained ground truth, affected by variation in the IRE, as $\kappa$ increases.
As a simple example, we numerically demonstrate that $f \in \{ f^0, f^1, f^2 \}$ minimizing $F(\kappa, f; w_0, f_0)$ at any $\kappa$ is not necessarily $f_0 \in \{ f^0, f^1, f^2 \}$ by using a nonlinear regression model with $K$ Gaussian components:
\begin{align}
f^K(x; w) = \sum_{k=1}^K a_k \exp \left(-\frac{b_k}{2} (x-c_k)^2 \right)
\label{eq:atomic}
\end{align}
for $K \geq 1$, where $w := \{a_k, b_k, c_k\}_{k=1}^K$ is the parameter set. This model is regarded as a kind of radial basis function networks \cite{broomhead1988multivariable}. 
Note that we define $f^0(x; w)=0$, where $w$ is an empty set. We also derive the analytic expression of $E(w;f, w_0, f_0)$ for this $f^K$ (see Appendix \ref{sec:App.D}).

Let us consider a {\it physical} interpretation of the {\it statistical ensemble} of nonlinear regression with $f^K$ as an analogy. Based on the mathematical equivalence between statistical inference and statistical physics, the {\it microstate} $w$ with the {\it particle number} $K$ subject to $p(w, f \mid  D^n, \beta, \mu)$ is interpreted as a {\it grand canonical ensemble} conditioned by $\beta$ as the {\it inverse temperature}, $n$ as the {\it volume}, and $\mu$ as the {\it chemical potential}. In the same manner, $w$ subject to $p(w \mid \beta, D^n, f)$ is interpreted as a {\it canonical ensemble}, where we assume $p(w, f \mid  D^n, \beta, \mu) = p(w \mid D^n, \beta, f) p(f \mid D^n, \beta, \mu)$ to derive Eq.\eqref{eq:ensemble}. Since both statistical ensembles are also conditioned by the {\it quenched disorder} $D^n$, we discuss the typical {\it macrostate} by means of the two averages $\langle \cdots \rangle$ and $[\cdots]$. For this purpose, $F(\kappa, f; w_0, f_0)$ and $\Lambda(\kappa, f; w_0, f_0)$ are reasonable in the limit where $\kappa$ and $L$ are fixed, whereas $n \to \infty$. It should be emphasized that we does not consider the {\it thermodynamic limit}, i.e., the limit where $K/n$ is fixed, whereas $K \to \infty$ and $n \to \infty$. 

We performed an Monte Carlo (MC) simulations by using parallel tempering based on the Metropolis criterion \cite{geyer1991markov, hukushima1996exchange}. The variable $\kappa$ was discretized as 400 points consisting of $\kappa=0$ and 399 logarithmically spaced points in the interval $[10^{-8}, 10^{12}]$. We set the prior probability distribution as $p(w \mid f^K) = \prod_{k=1}^K \exp (-a_k/10 - b_k/10 - c_k^2/50)/(500 \sqrt{2\pi})$. 
We simulated $p(w \mid D^n, \beta, f) \propto \exp \left( -\kappa E_n(w;f)/2 \right) p(w \mid f)$ with a realization of $D^n$ to calculate $F_n(\beta, f)$ in the same manner as in our previous work \cite{tokuda2017simultaneous}.
We also simulated $p(w \mid D^n, \beta, f) \propto \exp ( -\kappa E(w;f, w_0, f_0) /2 ) p(w \mid f)$ to calculate $F(\kappa, f; w_0, f_0)$ and $\Lambda(\kappa, f; w_0, f_0)$ at each point of $\kappa>0$ in the manner of Bridge sampling \cite{meng1996simulating, gelman1998simulating}. In all the MC simulations, the total MC sweeps were 100,000 after the burn-in. The error bars of $F(\kappa, f; w_0, f_0)$ and $\Lambda(\kappa, f; w_0, f_0)$ were calculated by bootstrap resampling. 

\begin{figure}[htbp]
\begin{center}
\begin{tabular}{c}
\includegraphics[width=8.6cm]{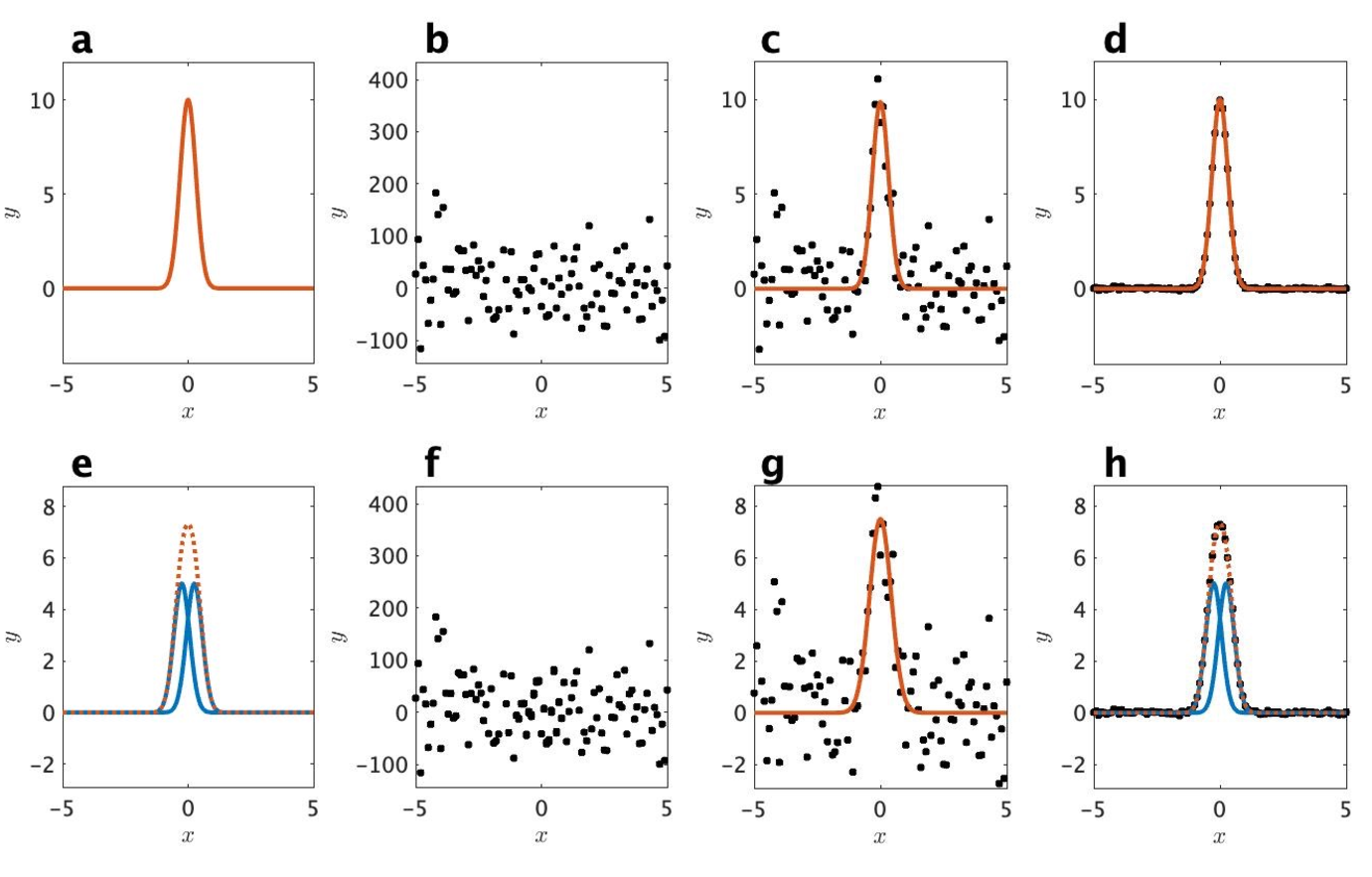}
\end{tabular}
\end{center}
\caption{Ground truth, observed data, and optimal model chosen by Bayesian model selection. It is demonstrated that which model minimizes $F_n(\beta, f)$ for each realization of $D^n$ (black dots) with $n=101$, $f=f^0(x;w)$ (no line), $f=f^1(x;w)$ (red solid line), or $f=f^2(x;w)$ (red dashed line) with two Gaussian components (blue solid lines).  
For the degenerate case that the ground truth is ({\bf a}) $f_0=f^1(x;w_0)$ with $w_0 = \{10, 10, 0\}$, the minimal model is ({\bf b}) $f=f^0$ at $\beta_0={\kappa}_1/n$, ({\bf c}) $f=f^1$ at $\beta_0={\kappa}_2/n$, and ({\bf d}) $f=f^1$ at $\beta_0={\kappa}_3/n$. 
For the splitting case that the ground truth is ({\bf e}) $f_0=f^2(x;w_0)$ with $w_0 = \{5, 10, (-1)^k \times 0.25 \}_{k=1}^2$, the minimal model is ({\bf f}) $f=f^0$ at $\beta_0={\kappa}_1/n$, ({\bf g}) $f=f^1$ at $\beta_0={\kappa}_2/n$, and ({\bf h}) $f=f^2$ at $\beta_0={\kappa}_3/n$.}
\label{fig:Fig.1}
\end{figure}

We consider two cases to simulate the discovery process of the gross and fine structures in observed data.
One is the degenerate case that is defined as $f_0 = f^1$ with $w_0= \{10, 10, 0\}$ (Fig. \ref{fig:Fig.1}a).
Another is the splitting case that is defined as $f_0 = f^2$ with $w_0=\{5, 10, (-1)^k \times 0.25 \}_{k=1}^{2}$, which makes two Gaussian components strongly overlapping (Fig. \ref{fig:Fig.1}e).
In each case, a realization of $D^n$ for $n=101$ is obtained in the presence of observation noise at a certain $\beta_0$. 
If $\beta_0$ is small enough, $f$ minimizing $F_n(\beta, f)$ in both cases are $f^0$, which is not consistent with $f_0$ (Figs. \ref{fig:Fig.1}b and \ref{fig:Fig.1}f).
If $\beta_0$ is large to some extent, $f$ minimizing $F_n(\beta, f)$ in both cases are $f^1$, which is consistent/inconsistent with $f_0$ in the degenerate/splitting case (Figs. \ref{fig:Fig.1}c and \ref{fig:Fig.1}g).
If $\beta_0$ is large enough, $f$ minimizing $F_n(\beta, f)$ in the splitting case is $f^2$, which is consistent with $f_0$ (Fig. \ref{fig:Fig.1}h).
They show that the optimal model chosen by the BMS is not always consistent with the ground truth and imply the typical behavior that the optimal model changes at some critical values of observations' fineness (rather than magnitude of observation noise);
For too rough observations, the optimal model just describes a "non-structure". For rather rough observations, the optimal model describes a "gross structure".
For fine observations, the optimal model describes a "fine structure".

\begin{figure*}[htbp]
\begin{center}
\begin{tabular}{c}
\includegraphics[width=8.6cm]{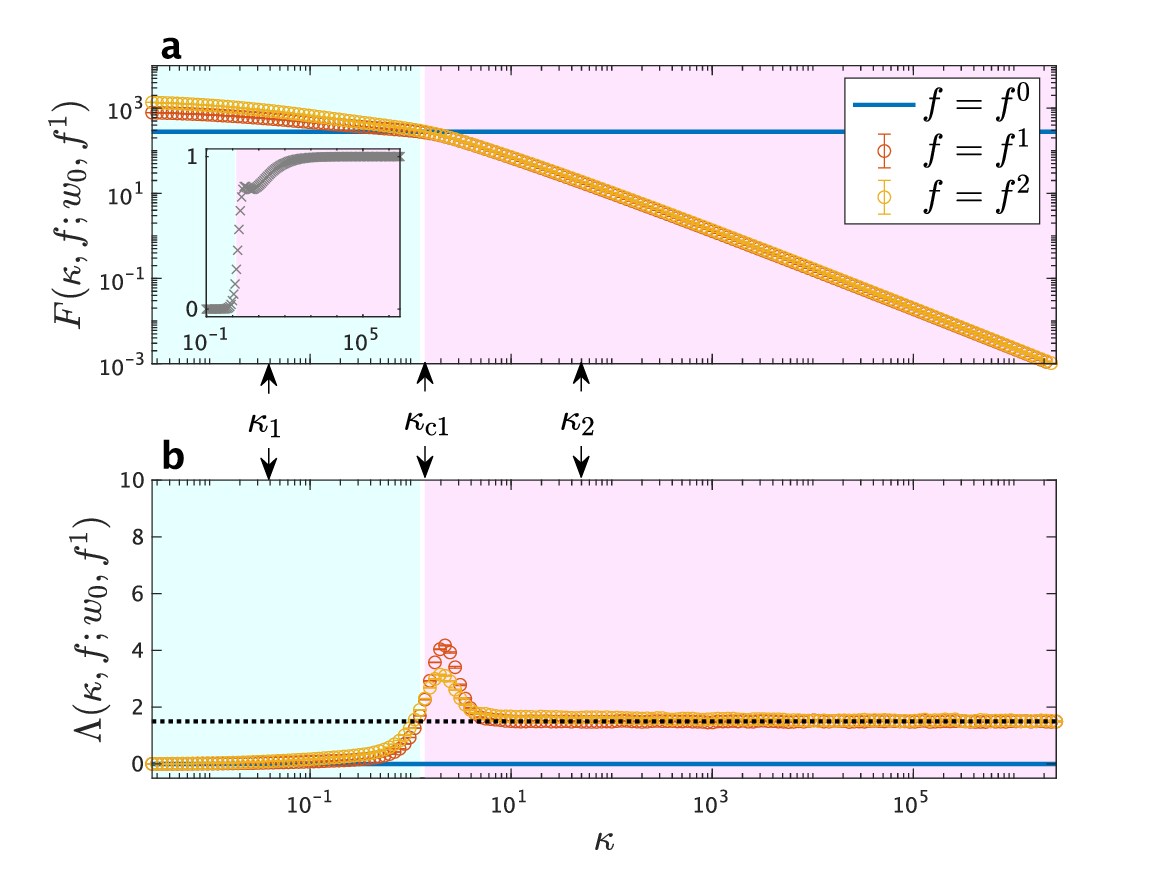}
\includegraphics[width=8.6cm]{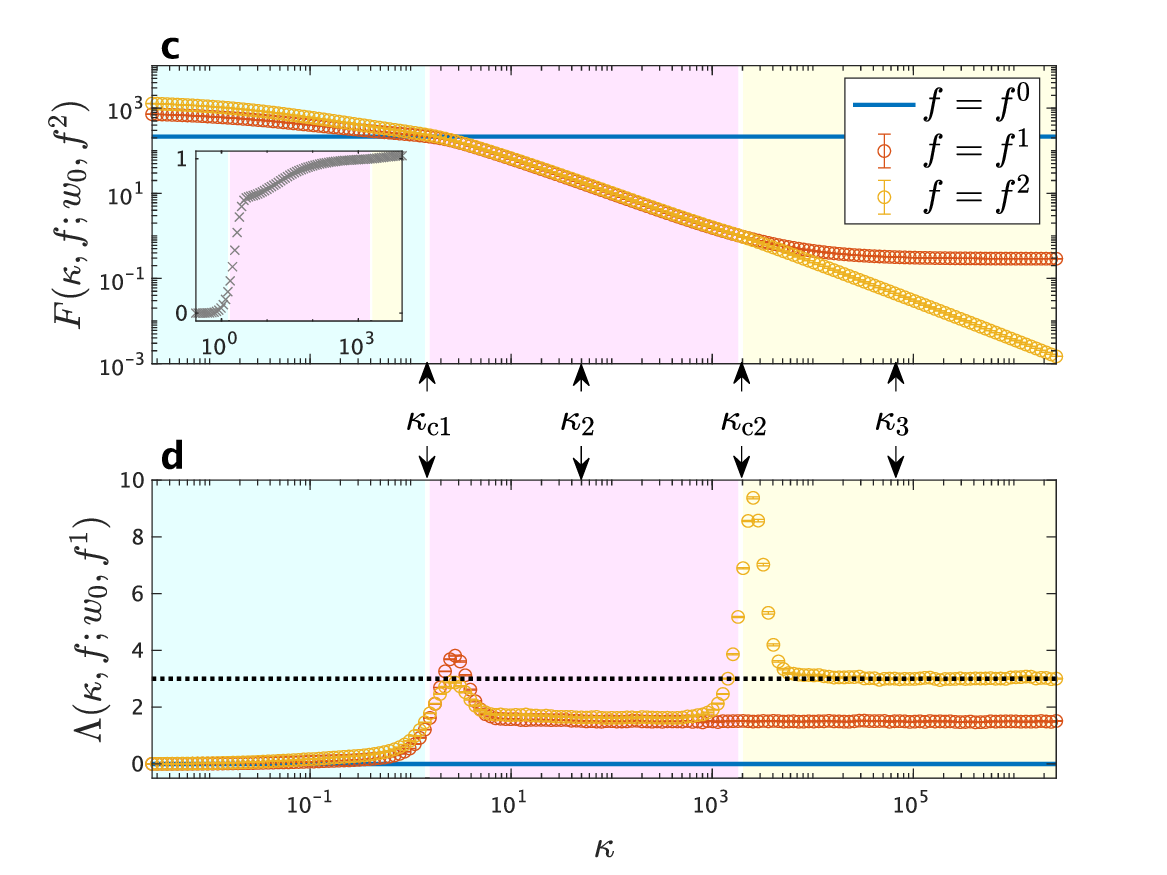}
\end{tabular}
\end{center}
\caption{Bayesian model selection and intrinsic regularization effect scaled by observations' fineness. The inequality $F(\kappa, f^0; w_0, f_0)<F(\kappa, f^1; w_0, f_0)$ holds for $\kappa<{\kappa}_{\rm c1}$ (region in light blue), $F(\kappa, f^0; w_0, f_0)>F(\kappa, f^1; w_0, f_0)$ holds for ${\kappa}_{\rm c1}<\kappa<{\kappa}_{\rm c2}$ (region in light pink), and $F(\kappa, f^1; w_0, f_0)>F(\kappa, f^2; w_0, f_0)$ holds for $\kappa>{\kappa}_{\rm c2}$ (region in light yellow). ({\bf a}) Log-log plot of $F(\kappa, f; w_0, f^1)$ for $w_0 = \{10, 10, 0\}$. 
Inset shows that the Bayes factor $\exp(-\kappa(F(\kappa, f^2; w_0, f^1)-F(\kappa, f^1; w_0, f^1)))<1$ holds for any $\kappa$. ({\bf b}) Semi-log plot of $\Lambda(\kappa, f; w_0, f^1)$ for $w_0 = \{10, 10, 0\}$ and asymptote ${\rm dim}(w)/2=1.5$ (black dotted line). ({\bf c}) Log-log plot of $F(\kappa, f; w_0, f^2)$ for $w_0 = \{5, 10, (-1)^k \times 0.25 \}_{k=1}^2$. 
Inset shows that the Bayes factor $\exp(-\kappa(F(\kappa, f^2; w_0, f^2)-F(\kappa, f^1; w_0, f^2)))<1$ holds for $\kappa<{\kappa}_{\rm c2}$, and $\exp(-\kappa(F(\kappa, f^2; w_0, f^2)-F(\kappa, f^1; w_0, f^2)))>1$ holds for $\kappa>{\kappa}_{\rm c2}$. ({\bf d}) Semi-log plot of $\Lambda(\kappa, f; w_0, f^2)$ for $w_0 = \{5, 10, (-1)^k \times 0.25 \}$ and asymptote ${\rm dim}(w)/2=3$ (black dotted line).}
\label{fig:Fig.2}
\end{figure*}

\begin{figure}[htbp]
\begin{center}
\begin{tabular}{c}
\includegraphics[width=8.6cm]{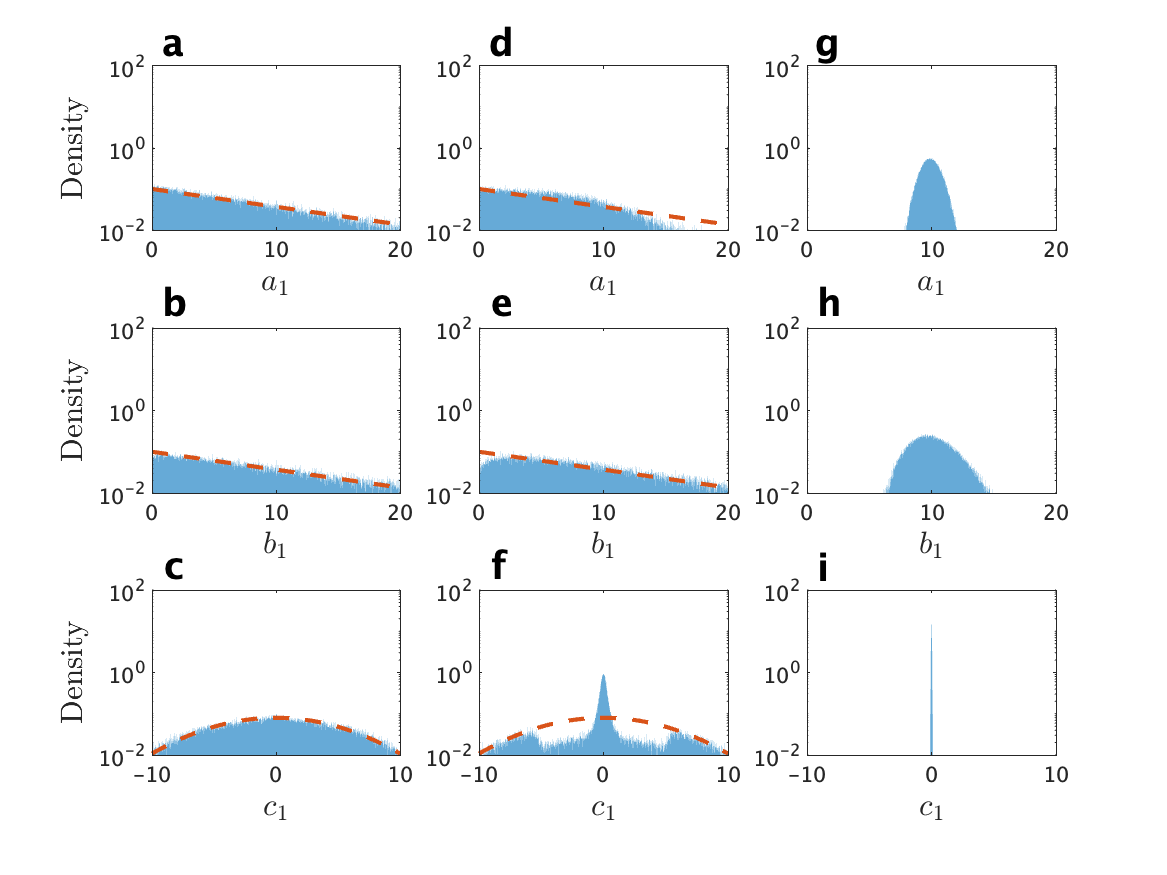}
\end{tabular}
\end{center}
\caption{Change in the posterior probability distribution from the prior probability distribution depending on observations' fineness. Histogram of the Monte Carlo sample from $p(w \mid D^n, \beta, f^1) \propto \exp ( -\kappa E(w;f^1, w_0, f^1) /2 ) p(w \mid f^1)$ for $w_0 = \{10, 10, 0\}$ at ({\bf a}-{\bf c}) $\kappa={\kappa}_1$, ({\bf d}-{\bf f}) $\kappa={\kappa}_{\rm c1}$, and ({\bf g}-{\bf i}) $\kappa={\kappa}_2$. Each row corresponds to a marginal distribution; ({\bf a}, {\bf d}, {\bf g}) corresponds to $p(a_1 \mid D^n, \beta, f^1)$, ({\bf b}, {\bf e}, {\bf h}) corresponds to $p(b_1 \mid D^n, \beta, f^1)$, and ({\bf c}, {\bf f}, {\bf i}) corresponds to $p(c_1 \mid D^n, \beta, f^1)$. Compare each histogram with the marginal distribution of $p(w \mid f^1)$ (red dashed line).}
\label{fig:Fig.3}
\end{figure}

To elucidate our implication, we calculate $F(\kappa, f; w_0, f_0)$ and $\Lambda(\kappa, f; w_0, f_0)$ on the Nishimori line $\kappa = n \beta_0$ for the above two cases.
In both cases above, the optimal model $f$ minimizing $F(\kappa, f; w_0, f_0)$ changes from $f^0$ to $f^1$ around $\kappa={\kappa}_{\rm c1}$ (Figs. \ref{fig:Fig.2}a and \ref{fig:Fig.2}c), while $\Lambda(\kappa, f^1; w_0, f_0)$ has a peak around $\kappa={\kappa}_{\rm c1}$ (Figs. \ref{fig:Fig.2}b and \ref{fig:Fig.2}d).
$\Lambda(\kappa, f^1; w_0, f_0)$ is fairly consistent with $0$ and $1.5$ at $\kappa \ll \kappa_{\rm c1}$ and $\kappa \gg \kappa_{\rm c1}$, respectively (Fig. \ref{fig:Fig.2}b). 
Corresponding changes in $p(w \mid D^n, \beta, f^1)$ are also shown (Fig. \ref{fig:Fig.3}).
While $p(w \mid D^n, \beta, f^1)$ at $\kappa=\kappa_1$ (Fig. \ref{fig:Fig.3}a-\ref{fig:Fig.3}c) is fairly consistent with $p(w \mid f^1)$, $p(w \mid D^n, \beta, f^1)$ at $\kappa=\kappa_2$ is sufficiently approximated by a Gaussian distribution whose mean is $w_0$ (Figs. \ref{fig:Fig.3}g-\ref{fig:Fig.3}i). $p(w \mid D^n, \beta, f^1)$ at $\kappa=\kappa_{\rm c1}$ represents the intermediate state (Figs. \ref{fig:Fig.3}d-\ref{fig:Fig.3}f) between these two states.

\begin{figure}[htbp]
\begin{center}
\begin{tabular}{c}
\includegraphics[width=8.6cm]{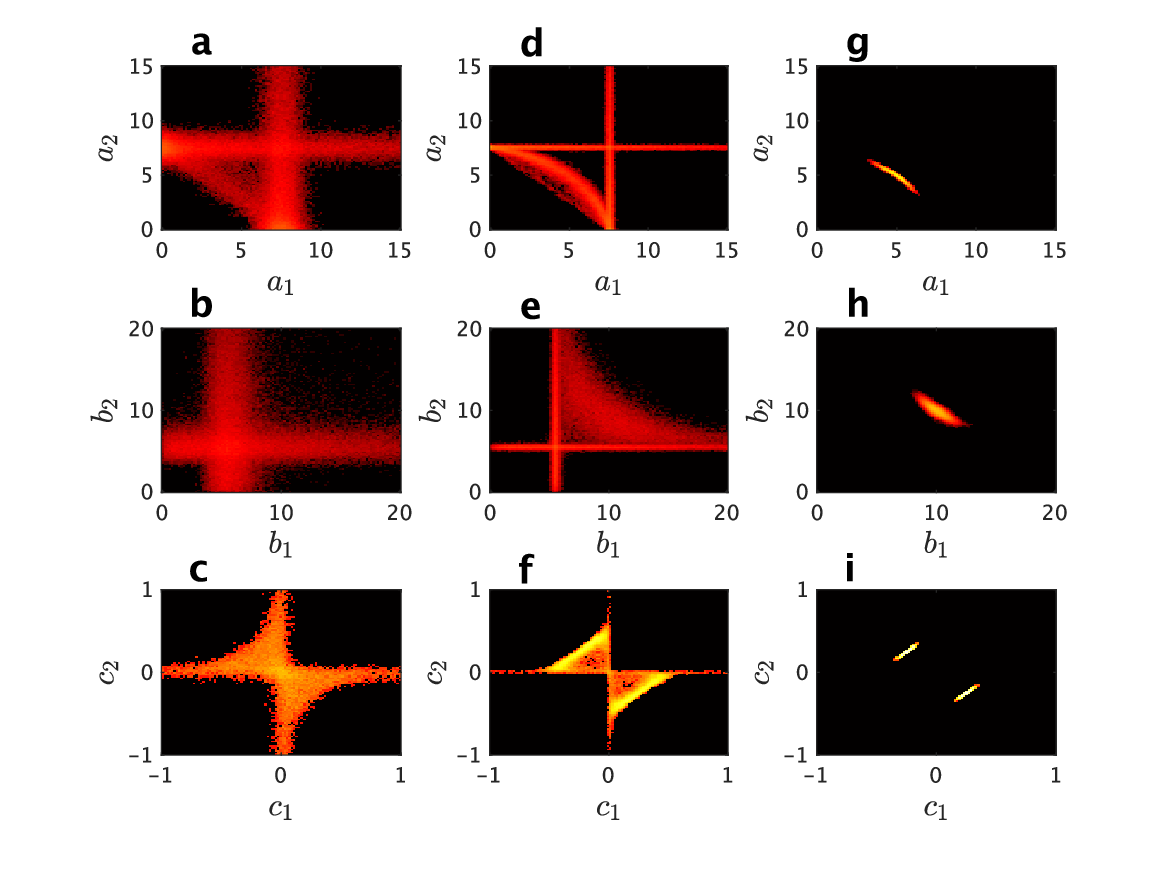}
\end{tabular}
\end{center}
\caption{Change in the posterior probability distribution depending on observations' fineness and underlying parameter correlations. Two-dimensional histogram of the Monte Carlo sample from $p(w \mid D^n, \beta, f^2) \propto \exp ( -\kappa E(w;f^2, w_0, f^2) /2 ) p(w \mid f^2)$ for $w_0 = \{5, 10, (-1)^k \times 0.25 \}_{k=1}^{2}$ at ({\bf a}-{\bf c}) $\kappa={\kappa}_2$, ({\bf d}-{\bf f}) $\kappa={\kappa}_{\rm c2}$, and ({\bf g}-{\bf i}) $\kappa={\kappa}_3$. Each row corresponds to a marginal distribution; ({\bf a}, {\bf d}, {\bf g}) corresponds to $p(a_1, a_2 \mid D^n, \beta, f^2)$, ({\bf b}, {\bf e}, {\bf h}) corresponds to $p(b_1, b_2 \mid D^n, \beta, f^2)$, and ({\bf c}, {\bf f}, {\bf i}) corresponds to $p(c_1, c_2 \mid D^n, \beta, f^2)$.}
\label{fig:Fig.4}
\end{figure}

Only in the splitting case, the optimal $f$ minimizing $F(\kappa, f; w_0, f_0)$ changes from $f^1$ to $f^2$ around $\kappa={\kappa}_{\rm c2}$ (Fig. \ref{fig:Fig.2}c), while $\Lambda(\kappa, f^1; w_0, f_0)$ has a peak around $\kappa={\kappa}_{\rm c2}$ (Fig. \ref{fig:Fig.2}d).
$\Lambda(\kappa, f^2; w_0, f_0)$ is fairly consistent with $1.5$ and $3$ at $\kappa_{\rm c1} \ll \kappa \ll \kappa_{\rm c2}$ and $\kappa \gg \kappa_{\rm c2}$, respectively (Fig. \ref{fig:Fig.2}d).
Corresponding changes in $p(w \mid D^n, \beta, f^2)$ are also shown (Fig. \ref{fig:Fig.4}).
While $p(w \mid D^n, \beta, f^2)$ at $\kappa=\kappa_2$ is far from a Gaussian distribution (Fig. \ref{fig:Fig.4}a-\ref{fig:Fig.4}c; see also Appendix \ref{sec:App.D}), $p(w \mid D^n, \beta, f^2)$ at $\kappa=\kappa_3$ is sufficiently approximated by a bimodal Gaussian distribution whose modes are symmetric (Fig. \ref{fig:Fig.4}g-\ref{fig:Fig.4}i). $p(w \mid D^n, \beta, f^2)$ at $\kappa=\kappa_{\rm c2}$ represents the intermediate state (Fig. \ref{fig:Fig.4}d-\ref{fig:Fig.4}f) between these two states.

\begin{figure*}[htbp]
\begin{center}
\begin{tabular}{c}
\includegraphics[width=8.6cm]{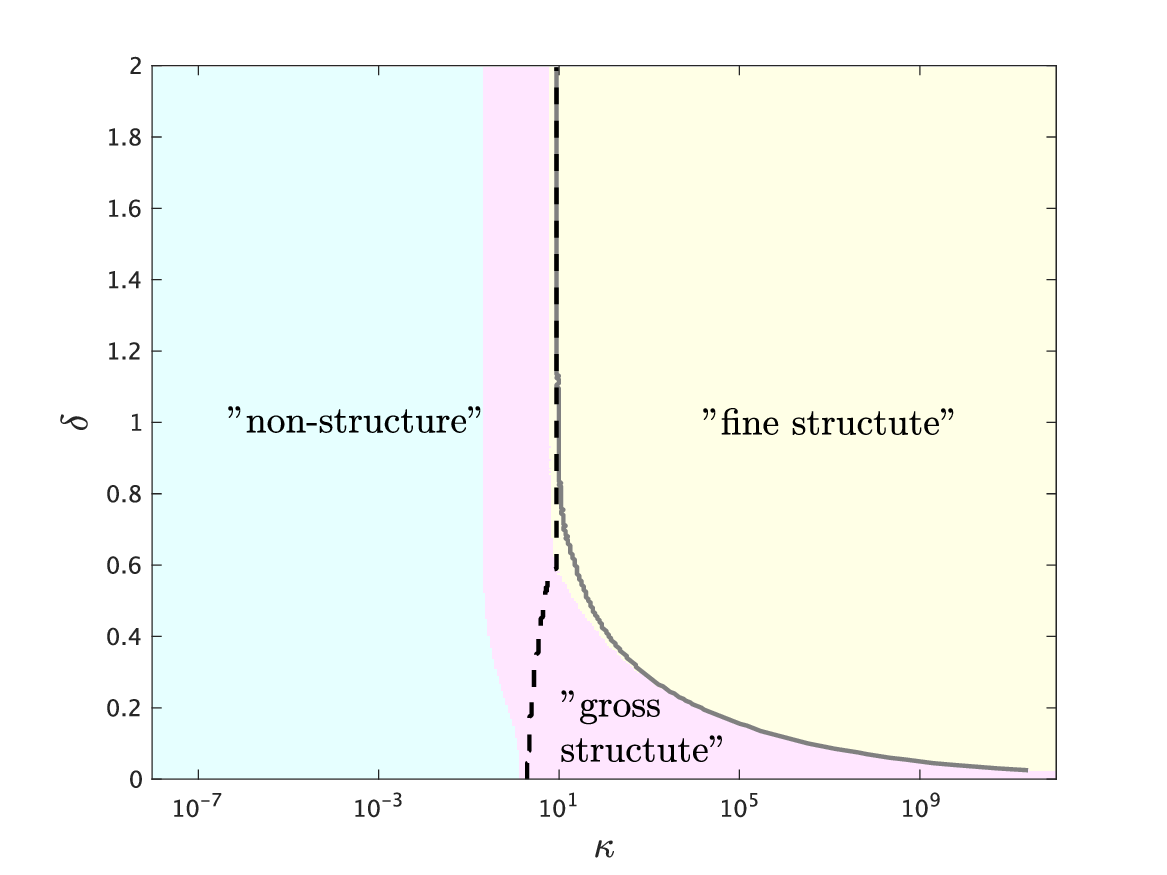}
\end{tabular}
\end{center}
\caption{Phase diagram with respect to observations' fineness and ground truth of  parameter. Semi-log plot of the peak positions of $\Lambda(\kappa, f^2; w_0, f^2)$  (black dashed and grey solid lines) for $w_0=\{5, 10, (-1)^k \times \delta \}_{k=1}^{2}$. The inequality $F(\kappa, f^0; w_0, f^2)<F(\kappa, f^1; w_0, f^2)$ holds for $\kappa<{\kappa}_{\rm c1}(\delta)$ (region in light blue), $F(\kappa, f^0; w_0, f^2)>F(\kappa, f^1; w_0, f^2)$ holds for ${\kappa}_{\rm c1}(\delta)<\kappa<{\kappa}_{\rm c2}(\delta)$ (region in light pink), and $F(\kappa, f^1; w_0, f^2)>F(\kappa, f^2; w_0, f^2)$ holds for $\kappa>{\kappa}_{\rm c2}(\delta)$ (region in light yellow).}
\label{fig:Fig.5}
\end{figure*}

We extend the investigation to the case that is defined as $f_0 = f^2$ with $w_0=\{5, 10, (-1)^k \times \delta \}_{k=1}^{2}$ for $0 \leq \delta \leq 2$, where $\delta=0$ and $\delta=0.25$ respectively correspond to the degenerate and splitting cases above. 
The {\it phase diagram} shows that there are three {\it phases} described by $\kappa$ and $\delta$ (Fig. \ref{fig:Fig.5}). 
Three phases are bounded by two ridge lines of $\Lambda(\kappa, f^2; w_0, f^2)$, where these lines merge in $\delta \gtrsim 0.8$. In other words, there is only two phases in $\delta \gtrsim 0.8$, while there are three phases $0 < \delta \lesssim 0.8$. Note that $\delta=0$ is a special case since the case of $f_0=f^2$ with $w_0= \{5, 10, 0\}_{k=1}^2$ is identified with the case of $f_0=f^1$ with $w_0= \{10, 10, 0\}$ (see Appendix \ref{sec:App.D}).  
The optimal model $f$ minimizing $F(\kappa, f; w_0, f_0)$ also changes around the {\it phase boundaries}. This correspondence enables us to interpret three phases as a "non-structure", a "gross structure", and a "fine structure". Note that the region in the "non-structure" phase, whereas the optimal model is $f=f^1$, corresponds to intermediate state, such as shown in Figs. \ref{fig:Fig.3}d-\ref{fig:Fig.3}f.

\section{Discussions}
Our results clearly show that the optimal model chosen by the BMS is not always consistent with the ground truth but depends on observations' fineness. The discovery process of the gross and fine structures shows the changes in the optimal model at critical values of observations' fineness, accompanying the variation in the IRE.

Our results can also be understood from another viewpoint.
An effective model is not necessary to be consistent with the optimal model for describing observed data since it is deduced from the original theory independent of the data. If one is more confident of an effective model than observed data, what model is optimal for observed data is replaced by what amount and quality of observed data are required to validate the model.
Our results show that critical values of observations' fineness are required to do so. From this point, these values can be regarded as kinds of limits on an indirect measurement, i.e., parameter estimation of the effective model.

\begin{acknowledgments}
The authors are grateful to Chihiro H. Nakajima, Koji Hukushima, Kouki Yonaga, Masayuki Ohzeki, Shotaro Akaho, Sumio Watanabe, Tomoyuki Obuchi and Yoshiyuki Kabashima for valuable discussions. S.T. was supported by JSPS KAKENHI (No. JP20K19889). M.O. was supported by JST CREST (No. JPMJCR1761), JSPS KAKENHI (No. 25120009), the "Materials Research by Information Integration" Initiative (MI2I) project of the Support Program for Starting Up Innovation Hub from the Japan Science and Technology Agency (JST), and the Council for Science, Technology and Innovation (CSTI), Cross-ministerial Strategic Innovation Promotion Program (SIP), "Structural Materials for Innovation" (Funding agency: JST).
\end{acknowledgments}




\appendix

\section{Definition of Bayes specific heat}
\label{sec:App.A}

Here, we derive Eq. \eqref{eq:SH} from a broader perspective of statistical inference including our setup. We start by introducing the conditional probability density
\begin{align}
p(w \mid D^n, \tilde{\beta}, \beta, f) :&= \frac{1}{\tilde{Z}_n(\tilde{\beta}, \beta, f)} \exp \left( - \frac{n \tilde{\beta}}{2} L_n(w; \beta, f) \right) p(w \mid f)
\label{eq:posterior_temp}
\end{align}
with $\tilde{\beta} \geq 0$ being the {\it inverse temperature} \cite{watanabe2009algebraic}, the empirical log loss function
\begin{align}
L_n(w; \beta, f) :&= - \frac{1}{n} \sum_{i=1}^n \log p(Y_i \mid x_i, w, \beta, f)
\end{align}
and the partition function
\begin{align}
\tilde{Z}_n(\tilde{\beta}, \beta, f) :&= \int \exp \left( - \frac{n \tilde{\beta}}{2} L_n(w; \beta, f) \right) p(w \mid f) dw.
\end{align}
Note that Eq. \eqref{eq:posterior_temp} for $\tilde{\beta}=1$ is just Bayes' theorem, where $p(w \mid D^n, 1, \beta, f)$ and $\tilde{Z}_n(1, \beta, f)$ are the posterior distribution and marginal likelihood, respectively. If $\tilde{\beta} \to \infty$ and $p(\hat{w} \mid f)>0$, then $p(w \mid D^n, \tilde{\beta}, \beta, f)$ converges to $\delta(w-\hat{w})$, where $\hat{w}$ is the maximum likelihood estimator \cite{watanabe2009algebraic}. Notably, $p(w \mid D^n, \tilde{\beta}, \beta, f)=p(w \mid f)$ holds for $\tilde{\beta}=0$.

Here, we define the specific heat
\begin{align}
\tilde{C}_n(\tilde{\beta}, \beta, f) :&= \frac{\partial \langle n L_n(w; \beta, f) \rangle_{\tilde{\beta}}}{\partial \tilde{\beta}^{-1}} = \tilde{\beta}^2 \tilde{I}_n(\tilde{\beta}; \beta, f)
\end{align}
with the Fisher information
\begin{align}
\tilde{I}_n(\tilde{\beta}; \beta , f) :&= \left \langle \left( \frac{\partial}{\partial \tilde{\beta}} \log p(w \mid D^n, \tilde{\beta}, \beta, f) \right)^2 \right \rangle_{\tilde{\beta}} = \langle (n L_n(w; \beta, f))^2 \rangle_{\tilde{\beta}} - \langle n L_n(w; \beta, f) \rangle_{\tilde{\beta}}^2,
\end{align}
where the average $\langle \cdots \rangle_{\tilde{\beta}} := \int ( \cdots ) p(w \mid D^n, \tilde{\beta}, \beta, f) dw$. 
Considering the connection between statistical inference and statistical physics, $\langle n L_n(w; \beta, f) \rangle_{\tilde{\beta}}$ is the internal energy and 
\begin{align}
\tilde{F}_n(\tilde{\beta}, \beta, f) :&= -\frac{1}{\tilde{\beta}} \log \tilde{Z}_n(\tilde{\beta}, \beta, f).
\end{align}
is the free energy. Then, we also obtain the relation
\begin{align}
\tilde{C}_n(\tilde{\beta}, \beta , f) &= - \tilde{\beta}^2 \frac{\partial^2 (\tilde{\beta} \tilde{F}_n)}{\partial \tilde{\beta}^2}
\label{eq:BSsec}
\end{align}
as in statistical physics. As the Bayes free energy is defined by $\tilde{F}_n(1, \beta, f)$, we define the Bayes specific heat as $\tilde{C}_n(1,\beta, f)$, where this definition can be applied not only to $L_n(w; \beta, f)$ in the nonlinear regression but also to empirical log loss functions of any other statistical inference setups without loss of generality. We should compare the Bayes specific heat, especially in the form of Eq. \eqref{eq:BSsec}, with the learning capacity \cite{lamont2019correspondence}, which is defined by the second derivative of the Bayes free energy with respect to $n$ as an approximation of the second-order-finite difference. Notably, the Bayes specific heat and the learning capacity are different, as $\tilde{\beta}$ and $n$ are different. However, they also have similarities. We show their similarities and differences in Appendix \ref{sec:App.C}.

Now, we consider the specifics of our setup, i.e., the relation
\begin{align}
L_n(w; \beta, f) &= \frac{\beta}{2} E_n(w; f) - \frac{1}{2} \log \frac{\beta}{2 \pi}.
\end{align}
Then, we obtain the scaling relations $\tilde{C}_n(\tilde{\beta}, \beta , f) = C_n( \tilde{\beta} \beta, f)$ and $\tilde{I}_n(\tilde{\beta}; \beta , f) = I_n(\tilde{\beta} \beta; f)$, where the scaling functions are
\begin{align}
C_n( \tilde{\beta} \beta, f) :&= (\tilde{\beta} \beta)^2 I_n(\tilde{\beta} \beta; f)
\end{align}
and
\begin{align}
I_n(\tilde{\beta} \beta; f):&= \langle (n E_n(w; f))^2 \rangle_{\tilde{\beta}} - \langle n E_n(w; f) \rangle_{\tilde{\beta}}^2.
\end{align}
Now, we take $\tilde{\beta}=1$, i.e $\tilde{\beta} \beta = \beta$, and then obtain $C_n(\beta, f)$ in the form of Eq. \eqref{eq:SH} as the Bayes specific heat.

\section{Derivation of scaling relation on Bayes free energy}
\label{sec:App.B}

Here, we show an outline of the derivation of Eqs. \eqref{eq:SFF} and \eqref{eq:SFF2}. By considering the noise additivity, we divided $Y_i$ into the signal and noise, i.e., 
\begin{align}
Y_i = f_0(x_i; w_0) + N_i,
\end{align}
where $N_i \sim \mathcal{N}(0, \beta_0^{-1})$. 
Then, we obtained
\begin{align}
n E_n(w;f) = \sum_{i} s_i(w;f)^2 + 2 \sum_{i} s_i(w;f) N_i + \frac{R_n}{\beta_0},
\label{eq:additivity}
\end{align}
where $s_i(w;f) := f_0(x_i;w_0) - f(x_i;w)$ and $R_n:=\beta_0\sum_{i} N_i^2$. By using Jensen's inequality,
\begin{align}
\left[- \log \int \exp \left( - \frac{\beta}{2} \left(\sum_{i} s_i(w;f)^2 + 2 \sum_{i} s_i(w;f) N_i \right) \right) p(w \mid f) dw \right] \geq - \log \int \exp \left( - \frac{\beta}{2} \sum_{i} s_i(w;f)^2 \right) p(w \mid f) dw
\end{align}
holds, where the equality holds when $\sum_{i} s_i(w;f) N_i = 0$, which is asymptotically satisfied for $n \to \infty$. Note that 
\begin{align}
\frac{1}{n} \sum_{i} s_i(w;f)^2 = E(w;f, w_0, f_0)
\end{align}
also holds as $n \to \infty$. Based on these asymptotic behaviors, Eqs. \eqref{eq:SFF} and \eqref{eq:SFF2} were obtained.

\section{Derivation and validation of scaling relation on Bayes specific heat}
\label{sec:App.C}
We show the derivation of Eqs. \eqref{eq:FSS2} and \eqref{eq:RLCT} in more detail. We start from a broader perspective of statistical inference including our setup. The asymptotic behaviour of the free energy has been obtained \cite{watanabe2009algebraic}: 
\begin{align}
\tilde{\beta} \tilde{F}_n(\tilde{\beta}, \beta, f) &= n \tilde{\beta} L_n(w_0'; \beta, f) + \lambda \log n \tilde{\beta} + (m-1) \log \log n \tilde{\beta} + O_p(\tilde{\beta})
\label{eq:asymptic_F}
\end{align}
as $n \to \infty$, where $w_0'$ is $w$ that minimizes the Kullback-Leibler distance from $p(y \mid x, w_0, \beta_0, f_0)$ to $p(y \mid x, w, \beta, f)$, $\lambda > 0$ is a rational number called the real log canonical threshold, and $m \geq 1$ is a natural number.  Note that $w_0' = w_0$ holds if $\beta=\beta_0$ and $f=f_0$. By following Eqs. \eqref{eq:BSsec} and \eqref{eq:asymptic_F}, we obtain
\begin{align}
\tilde{C}_n(\tilde{\beta}, \beta, f) &=  \lambda - (m-1) \left( \frac{1}{ \log n \tilde{\beta}} + \frac{1}{( \log n \tilde{\beta} )^2} \right) + o_p \left( \tilde{\beta}^2 \right)
\label{eq:scaling_anal}
\end{align}
as $n \to \infty$. If we take $\tilde{\beta} \to \infty$, then $\tilde{C}_n(\tilde{\beta}, \beta, f) =  \lambda + o_p \left( \tilde{\beta}^2 \right)$ holds; the quantity $\tilde{C}_n(\tilde{\beta}, \beta, f)$ is not necessarily self-averaging. The relation $\tilde{C}_n(\tilde{\beta}, \beta, f) = \lambda$ holds as $n \to \infty$ if $\tilde{\beta}=O(1/\log n)$, which corresponds to the condition shown in Watanabe's corollary \cite{watanabe2013widely}:
\begin{align}
\frac{\langle n L_n(w; \beta, f) \rangle_{\tilde{\beta}_1} - \langle n L_n(w; \beta, f) \rangle_{\tilde{\beta}_2}}{\tilde{\beta}_1^{-1}-\tilde{\beta}_2^{-1}} = \lambda + O_p \left( \frac{1}{\sqrt{\log n}} \right)
\end{align}
hold for $\tilde{\beta}_1=O(1/\log n)$ and $\tilde{\beta}_2=O(1/\log n)$, where $\tilde{\beta}_1$ and $\tilde{\beta}_2$ are positive variables. 
Therefore, it is recertified that $\tilde{C}_n(\tilde{\beta}, \beta, f) = \lambda$ holds for $n \to \infty$ if $\tilde{\beta}=O(1/\log n)$, where
\begin{align}
\frac{\langle n L_n(w; \beta, f) \rangle_{\tilde{\beta}_1} - \langle n L_n(w; \beta, f) \rangle_{\tilde{\beta}_2}}{\tilde{\beta}_1^{-1}-\tilde{\beta}_2^{-1}} = \tilde{C}_n(\tilde{\beta}, \beta, f)
\end{align}
hold as $\tilde{\beta}_1 \to \tilde{\beta}$ and $\tilde{\beta}_2 \to \tilde{\beta}$.
Here, we mention that the expectation of the learning capacity over realizations also converges to $\lambda$ as $n \to \infty$ \cite{lamont2019correspondence}.
However, it has not proven that the learning capacity as a random variable converges toward $\lambda$ as $n \to \infty$.  
The learning capacity as a random variable is not applicable for the scaling analysis that provides Eq. \eqref{eq:scaling_anal}. It does not also provide Eqs. \eqref{eq:FSS2} and \eqref{eq:RLCT} in the limit where $\kappa$ and $L$ are fixed, whereas $n \to \infty$.

Now, we consider the specifics of our setup, i.e., the scaling relation
\begin{align}
C_n( \tilde{\beta} \beta, f) &=  \lambda - (m-1) \left( \frac{1}{ \log \tilde{\beta} \kappa} + \frac{1}{( \log \tilde{\beta} \kappa )^2} \right) + o_p \left( {\tilde{\beta}^2 \beta}^2 \right),
\end{align}
where the correspondence of $(\tilde{\beta}, \beta)$ and $\tilde{\beta} \beta$ is considered. Then, we obtain 
\begin{align}
C_n(\beta, f) &=  \lambda - (m-1) \left( \frac{1}{ \log \kappa} + \frac{1}{( \log \kappa)^2} \right) + o_p \left( \beta^2 \right)
\label{eq:SH1}
\end{align}
for $\tilde{\beta}=1$.

We also evaluate the term $o_p \left( \beta^2 \right)$ more tightly. Following Eqs. (2) and \eqref{eq:additivity}, we obtain
\begin{align}
C_n(\beta, f) &= \left( \frac{\beta}{2} \right)^2 \left( \left \langle \left( \sum_{i=1}^{n} s_i(w; f)^2 \right)^2 \right \rangle - \left \langle \sum_{i=1}^{n}  s_i(w; f)^2 \right \rangle^2 \right) +  V_n(\beta, f) + \tilde{V}_n(\beta, f) + W_n(\beta, f) + \tilde{W}_n(\beta, f),
\label{eq:SHtot}
\end{align}
where
\begin{align}
V_n(\beta, f) :&= \beta^2 \sum_{i \neq j} \left( \left \langle s_i(w; f) s_j(w; f) \right \rangle - \left \langle s_i(w; f) \right \rangle \left \langle s_j(w; f) \right \rangle \right) N_i N_j,
\end{align}
\begin{align}
\tilde{V}_n(\beta, f) :&= \beta^2 \sum_{i=1}^n \left( \left \langle s_i(w; f)^2 \right \rangle - \left \langle s_i(w; f) \right \rangle^2 \right) N_i^2,
\end{align}
\begin{align}
W_n(\beta, f) :&= \beta^2 \sum_{i \neq j} \left( \left \langle s_i(w; f)^2 s_j(w; f) \right \rangle - \left \langle s_i(w; f)^2 \right \rangle \left \langle s_j(w; f) \right \rangle \right) N_j,
\end{align}
and
\begin{align}
\tilde{W}_n(\beta, f) := \beta^2 \sum_{i=1}^n \left( \left \langle s_i(w; f)^3 \right \rangle - \left \langle s_i(w; f)^2 \right \rangle \left \langle s_i(w; f) \right \rangle \right) N_i.
\end{align}

We evaluate the order of each term in Eq. \eqref{eq:SHtot} in the limit where $\kappa$ and $L$ are fixed, whereas $n \to \infty$. First, we obtain
\begin{align}
\left( \frac{\beta}{2} \right)^2 \left( \left \langle \left( \sum_{i=1}^{n} s_i(w; f)^2 \right)^2 \right \rangle - \left \langle \sum_{i=1}^{n}  s_i(w; f)^2 \right \rangle^2 \right) &= \Lambda(\kappa, f; w_0, f_0)
\label{eq:Lambda}
\end{align}
in the limit that we consider.

Second, we evaluate the order of $V_n(\beta, f)$ as $n \to \infty$. Now, we obtain 
\begin{align}
\left[ V_n(\beta, f) \right] &=0
\end{align}
as $n \to \infty$, such that
\begin{align}
[\langle \left( \left \langle s_i(w; f) s_j(w; f) \right \rangle - \left \langle s_i(w; f) \right \rangle \left \langle s_j(w; f) \right \rangle \rangle \right) N_i N_j] = \left( \left \langle s_i(w; f) s_j(w; f) \right \rangle - \left \langle s_i(w; f) \right \rangle \left \langle s_j(w; f) \right \rangle \right) [N_i] [N_j]
\end{align}
is satisfied, where $[N_i]=0$. Then, we also obtain
\begin{align}
\left[ V_n(\beta, f)^2\right] - \left[ V_n(\beta, f) \right]^2 &= \frac{\beta^4}{\beta_0^2} \sum_{i \neq j} \left( \left \langle s_i(w; f) s_j(w; f) \right \rangle - \left \langle s_i(w; f) \right \rangle \left \langle s_j(w; f) \right \rangle \right)^2 \notag \\
&= O \left( \frac{\beta^2}{\beta_0^2} \right)
\end{align}
as $n \to \infty$, such that
\begin{align}
\left[ \left( \left \langle s_i(w; f) s_j(w; f) \right \rangle - \left \langle s_i(w; f) \right \rangle \left \langle s_j(w; f) \right \rangle \right)^2 N_i^2 N_j^2 \right] &=
\left( \left \langle s_i(w; f) s_j(w; f) \right \rangle - \left \langle s_i(w; f) \right \rangle \left \langle s_j(w; f) \right \rangle \right)^2 [N_i^2] [N_j^2]
\end{align}
and
\begin{align}
\left[ \left( \left \langle s_i(w; f) s_j(w; f) \right \rangle - \left \langle s_i(w; f) \right \rangle \left \langle s_j(w; f) \right \rangle \right) \left( \left \langle s_k(w; f) s_l(w; f) \right \rangle - \left \langle s_k(w; f) \right \rangle \left \langle s_l(w; f) \right \rangle \right)
N_i N_j N_k N_l \right] \notag \\
= \left( \left \langle s_i(w; f) s_j(w; f) \right \rangle - \left \langle s_i(w; f) \right \rangle \left \langle s_j(w; f) \right \rangle \right) \left( \left \langle s_k(w; f) s_l(w; f) \right \rangle - \left \langle s_k(w; f) \right \rangle \left \langle s_l(w; f) \right \rangle \right)
[N_i] [N_j] [N_k] [N_l]
\end{align}
are satisfied, where $[N_i]=0$, $[N_i^2]=\beta_0^{-1}$, $\left \langle s_i(w; f) \right \rangle = O(\kappa^{-1})$ and $\left \langle s_i(w; f) s_j(w; f) \right \rangle = O(\kappa^{-1})$. In summary, we obtain
\begin{align}
V_n(\beta, f) &= \left[ V_n(\beta, f) \right] + O_p \left( \frac{\beta}{\beta_0} \right) \notag \\
&=O_p \left( \frac{\beta}{\beta_0} \right)
\label{eq:Ord1}
\end{align}
as $n \to \infty$.

In the same way, we also obtain
\begin{align}
\tilde{V}_n(\beta, f) &= \left[ \tilde{V}_n(\beta, f) \right] + O_p \left( \frac{\beta}{\sqrt{n} \beta_0} \right)
\label{eq:Ord2}
\end{align}
as $n \to \infty$ with
\begin{align}
\left[ \tilde{V}_n(\beta, f) \right] &= \frac{\beta^2}{\beta_0} \sum_{i=1}^n \left( \left \langle s_i(w; f)^2 \right \rangle - \left \langle s_i(w; f) \right \rangle^2 \right) \notag \\
&= O \left( \frac{\beta}{\beta_0} \right),
\end{align}
where $[N_i^2]=\beta_0^{-1}$, $\left[ N_i^4 \right] = 3/\beta_0^2$, $\left \langle s_i(w; f) \right \rangle = O(\kappa^{-1})$ and
$\left \langle s_i(w; f)^2 \right \rangle = O(\kappa^{-1})$.
This means that $\tilde{V}_n$ is self-averaging; i.e., $\tilde{V}_n = [\tilde{V}_n] \geq 0$ holds as $n \to \infty$.
Furthermore, we also obtain
\begin{align}
W_n(\beta, f) &= O_p \left( \frac{1}{n \sqrt{\beta_0}} \right),
\end{align}
and
\begin{align}
\tilde{W}_n(\beta, f) &= O_p \left( \frac{1}{n^{3/2} \sqrt{\beta_0}} \right)
\label{eq:Ord4}
\end{align}
as $n \to \infty$, where $[N_i]=0$, $[N_i^2]=\beta_0^{-1}$, $[W_n]=0$, $[\tilde{W}_n]=0$,
$\left \langle s_i(w; f) \right \rangle = O(\kappa^{-1})$,
$\left \langle s_i(w; f)^2 \right \rangle = O(\kappa^{-1})$,
$\left \langle s_i(w; f)^2 s_j(w; f) \right \rangle = O \left( \kappa^{-2} \right)$,
and $\left \langle s_i(w; f)^3 \right \rangle = O \left( \kappa^{-2} \right)$.
By considering the consistency between Eqs. \eqref{eq:SH1} and \eqref{eq:SHtot} with Eqs. \eqref{eq:Lambda} and (\ref{eq:Ord1}-\ref{eq:Ord4}), we obtain Eqs. \eqref{eq:FSS2} and \eqref{eq:RLCT}, where $\Lambda(\kappa, f; w_0, f_0)=\lambda$ as $\kappa \to \infty$ is the real log canonical threshold. From this asymptotic behavior and Eq. \eqref{eq:asymptic_F}, it is found out that $\Lambda(\kappa, f; w_0, f_0)$ and $C_n(\beta, f)$ represent the intrinsic regularization effect in $F_n(\beta, f)$ as $\kappa \to \infty$:
\begin{align}
\kappa F_n(\beta, f) &= n L_n(w_0'; \beta, f) + \lambda \log n + (m-1) \log \log n + \frac{n}{2} \log \frac{\beta}{2 \pi}+ O_p (1),
\label{eq:leading}
\end{align}
where $\tilde{F}_n(1, \beta, f) = \kappa F_n(\beta, f) - n (\log \beta -\log 2 \pi)/2$  holds \cite{tokuda2017simultaneous}.

Here, we demonstrate the validity of Eqs. \eqref{eq:FSS2} and \eqref{eq:RLCT}. By performing the same simulation in Fig. 1 based on 100 different realizations of $D^n$ for $n=101$ taken from identical $p(y_i \mid x_i, w_0, \beta_0)$, we calculate $C_n(\beta, f)$ for each realization. At any $\beta$, the expectation of $C_n(\beta, f^1)$ over the realizations is fairly consistent with $\Lambda(\kappa, f^1; w_0, f^1)$ if $\beta_0=\beta_2$ (Fig. \ref{fig:Fig.S1}a), $\Lambda(\kappa, f^1; w_0, f^2)$ (Fig. \ref{fig:Fig.S1}c), and $\Lambda(\kappa, f^2; w_0, f^2)$ if $\beta_0=\beta_3$.
The standard deviations of $C_n(\beta, f)$ for the realizations is small enough (Fig. \ref{fig:Fig.S2}); the quantity $C_n(\beta, f)$ is considered to be self-averaging at any $\beta$ in these cases without the condition $\beta=O(\beta_0/\log n)$. According to Eq. \eqref{eq:FSS2}, these cases mean that the expectation of $C_n(\beta, f)$ corresponds to the average term $\Lambda(\kappa, f; w_0, f_0)$, where the standard deviation of $C_n(\beta, f)$ corresponds to the fluctuation term of order $O_p \left( \max \left( (\log \kappa)^{-1}, \beta/\beta_0, (n \sqrt{\beta_0})^{-1} \right) \right)$. Note that the standard deviation of $C_n(\beta, f)$ as the fluctuation term shows a dependence on $\Lambda(\kappa, f; w_0, f_0)$ (Fig. \ref{fig:Fig.S2}), which is not predicted by Eq. \eqref{eq:FSS2}. 

In other cases, the expectation of $C_n(\beta, f)$ is not consistent with $\Lambda(\kappa, f; w_0, f_0)$ at $\beta \gtrsim \beta_0$ (Fig. \ref{fig:Fig.S1}); finite-size effects on $C_n(\beta, f)$ appear at $\beta \gtrsim \beta_0$, where the term of order $O_p(\beta/\beta_0)$ is dominant.
According to Eq. \eqref{eq:SHtot} with Eqs. (\ref{eq:Ord1}-\ref{eq:Ord4}), the expectation of $C_n(\beta, f)$ corresponds to $\Lambda(\kappa, f; w_0, f_0)+\tilde{V}_n(\beta, f)$ as $n \to \infty$, where the standard deviation of $C_n(\beta, f)$ corresponds to $V_n(\beta, f) + W_n(\beta, f) + \tilde{W}_n(\beta, f)$ as $n \to \infty$. The expectations and standard deviations of $C_n(\beta, f)$ are roughly proportional to $\sqrt{\beta}$ if $\beta$ is large enough (Fig. \ref{fig:Fig.S3}) and also show a rough dependence on $\beta_0^2$. 
Note that $C_n(\beta, f)$ is considered to be self-averaging for $\beta=O(\beta_0/\log n)$ in all cases.

\section{Energy function of radial basis function network}
\label{sec:App.D}

Here, we derive the analytic expression of $E(w;f, w_0, f_0)$ for $f, f_0 \in \{ f^K\}$, where $f^K$ is defined as Eq. \eqref{eq:atomic}: 
\begin{align}
E(w;f, w_0, f_0) &= H(w_0, w_0) - 2 H(w_0, w) + H(w, w)
\end{align}
with
\begin{align}
H(w, w') :&= \frac{1}{L}\sum_{j=1}^{K} \sum_{k=1}^{K'} a_j a'_k \sqrt{\frac{2}{ b_j + {b'}_k}} \exp \left( - \frac{\left(c_j - c'_k \right)^2}{2 \left( b_j^{-1} + b_k'^{-1} \right)} \right)  \tilde{H}(b_j, c_j, {b'}_k, c'_k)
\end{align}
for $w' := \{a'_k, {b'}_k, c'_k\}_{k=1}^{K'}$ as $K K' > 0$ and $H(w, w'):=0$ as $K K' = 0$, where
\begin{align}
\tilde{H}(b_j, c_j, {b'}_k, c'_k) :&= \frac{\sqrt{\pi}}{2} {\rm erf} \left( \sqrt{\frac{b_j+{b'}_k}{2}} \left( x_n - \frac{c_j b_j + c'_k {b'}_k }{b_j+{b'}_k} \right) \right) - \frac{\sqrt{\pi}}{2} {\rm erf} \left( \sqrt{\frac{b_j+{b'}_k}{2}} \left( x_1 - \frac{c_j b_j + c'_k {b'}_k}{b_j + {b'}_k} \right) \right).
\end{align}
Note that $\tilde{H}(b_j, c_j, {b'}_k, c'_k) = \sqrt{\pi}$ holds as $x_1 \to - \infty$ and $x_n \to \infty$. 

We explain why $p(w \mid D^n, \beta, f^2)$ at $\kappa={\kappa}_2$ (Figs.4a-4c) exhibits such a correlation in the parameters.
Although the ground truth are $f_0=f^2$ with $w_0= \{5, 10, (-1)^k \times 0.25 \}_{k=1}^{2}$, here, we consider the pseudo-ground truth $\tilde{f}_0=f^1$ with $\tilde{w}_0:= \{ a^*, b^*, c^* \}$ and the analytic set 
\begin{align}
\tilde{W}_0 := \{ w \mid f(x_i;w) = \tilde{f}_0(x_i;\tilde{w}_0) \} = \tilde{W}_{01}  \cup \tilde{W}_{02} \cup \tilde{W}_{03}
\end{align}
for $w := \{a_k, b_k, c_k\}_{k=1}^2$, where
\begin{align}
\tilde{W}_{01} := \{ w \mid a_k + a_{k'} = a^*, b_k = b_{k'} = b^*, c_k = c_{k'}= c^* \}, \\
\tilde{W}_{02} := \{ w \mid a_k= a^*, a_{k'} =0 , b_k = b^*, c_k = c^* \},
\end{align}
and
\begin{align}
\tilde{W}_{03}:= \left\{ w \mid a_k= a^*, b_k = b^*, c_k = c^*, \exp \left(-\frac{b_{k'}}{2} (x_i-c_{k'})^2 \right) = 0 \right\}
\end{align}
for $i=1, \cdots, n$ and $k \neq k'$. Note that $\exp (-b_{k'}(x_i-c_{k'})^2 /2 )=0$ holds as $b_{k'} \to \infty$ or $c_{k'} \to \pm \infty$, where $b_{k'} \neq 0$ and $c_{k'} \neq x_i$ are necessary. Notably, $E(\tilde{w};f^2, \tilde{w}_0, \tilde{f}_0)=0$ holds for $\forall \tilde{w} \in \tilde{W}_0$; i.e., $p(w \mid D^n, \beta, f^2)$ can be relatively large around $w=\tilde{w}$ for $p (w \mid f^2) > 0$. The above scenario of the pseudo-ground truth corresponds to $p(w \mid D^n, \beta, f^2)$ at $\kappa={\kappa}_2$ (Figs. 4a-4c); i.e., $w \simeq \tilde{w}_0$ is statistically optimal for $f=f^2$ at $\kappa={\kappa}_2$. This scenario qualitatively holds at any ${\kappa}_{\rm c1} \ll \kappa \ll {\kappa}_{\rm c2}$, since $\Lambda(\kappa, f^2; w_0, f^2)$ is fairly consistent with $1.5(=\Lambda(\kappa, f^1; w_0, f^2))$.
Note that this result is from the situation where $f_0=f^2(x;w_0)$ appears almost as one Gaussian component (Fig. 1e), where one can only capture a "gross structure" depending on observations' fineness. 


\begin{thebibliography}{43}%
\makeatletter
\providecommand \@ifxundefined [1]{%
 \@ifx{#1\undefined}
}%
\providecommand \@ifnum [1]{%
 \ifnum #1\expandafter \@firstoftwo
 \else \expandafter \@secondoftwo
 \fi
}%
\providecommand \@ifx [1]{%
 \ifx #1\expandafter \@firstoftwo
 \else \expandafter \@secondoftwo
 \fi
}%
\providecommand \natexlab [1]{#1}%
\providecommand \enquote  [1]{``#1''}%
\providecommand \bibnamefont  [1]{#1}%
\providecommand \bibfnamefont [1]{#1}%
\providecommand \citenamefont [1]{#1}%
\providecommand \href@noop [0]{\@secondoftwo}%
\providecommand \href [0]{\begingroup \@sanitize@url \@href}%
\providecommand \@href[1]{\@@startlink{#1}\@@href}%
\providecommand \@@href[1]{\endgroup#1\@@endlink}%
\providecommand \@sanitize@url [0]{\catcode `\\12\catcode `\$12\catcode
  `\&12\catcode `\#12\catcode `\^12\catcode `\_12\catcode `\%12\relax}%
\providecommand \@@startlink[1]{}%
\providecommand \@@endlink[0]{}%
\providecommand \url  [0]{\begingroup\@sanitize@url \@url }%
\providecommand \@url [1]{\endgroup\@href {#1}{\urlprefix }}%
\providecommand \urlprefix  [0]{URL }%
\providecommand \Eprint [0]{\href }%
\providecommand \doibase [0]{https://doi.org/}%
\providecommand \selectlanguage [0]{\@gobble}%
\providecommand \bibinfo  [0]{\@secondoftwo}%
\providecommand \bibfield  [0]{\@secondoftwo}%
\providecommand \translation [1]{[#1]}%
\providecommand \BibitemOpen [0]{}%
\providecommand \bibitemStop [0]{}%
\providecommand \bibitemNoStop [0]{.\EOS\space}%
\providecommand \EOS [0]{\spacefactor3000\relax}%
\providecommand \BibitemShut  [1]{\csname bibitem#1\endcsname}%
\let\auto@bib@innerbib\@empty
\bibitem [{\citenamefont {Goldenfeld}(1992)}]{goldenfeld1992lectures}%
  \BibitemOpen
  \bibfield  {author} {\bibinfo {author} {\bibfnamefont {N.~D.}\ \bibnamefont
  {Goldenfeld}},\ }\href@noop {} {\emph {\bibinfo {title} {Lectures on phase
  transitions and the renormalization group}}}\ (\bibinfo  {publisher}
  {Addison-Wesley},\ \bibinfo {year} {1992})\BibitemShut {NoStop}%
\bibitem [{\citenamefont {Goldenfeld}\ and\ \citenamefont
  {Kadanoff}(1999)}]{goldenfeld1999simple}%
  \BibitemOpen
  \bibfield  {author} {\bibinfo {author} {\bibfnamefont {N.}~\bibnamefont
  {Goldenfeld}}\ and\ \bibinfo {author} {\bibfnamefont {L.~P.}\ \bibnamefont
  {Kadanoff}},\ }\bibfield  {title} {\bibinfo {title} {Simple lessons from
  complexity},\ }\href@noop {} {\bibfield  {journal} {\bibinfo  {journal}
  {Science}\ }\textbf {\bibinfo {volume} {284}},\ \bibinfo {pages} {87}
  (\bibinfo {year} {1999})}\BibitemShut {NoStop}%
\bibitem [{\citenamefont {Batterman}(2002)}]{batterman2002asymptotics}%
  \BibitemOpen
  \bibfield  {author} {\bibinfo {author} {\bibfnamefont {R.~W.}\ \bibnamefont
  {Batterman}},\ }\bibfield  {title} {\bibinfo {title} {Asymptotics and the
  role of minimal models},\ }\href@noop {} {\bibfield  {journal} {\bibinfo
  {journal} {The British Journal for the Philosophy of Science}\ }\textbf
  {\bibinfo {volume} {53}},\ \bibinfo {pages} {21} (\bibinfo {year}
  {2002})}\BibitemShut {NoStop}%
\bibitem [{\citenamefont {Batterman}\ and\ \citenamefont
  {Rice}(2014)}]{batterman2014minimal}%
  \BibitemOpen
  \bibfield  {author} {\bibinfo {author} {\bibfnamefont {R.~W.}\ \bibnamefont
  {Batterman}}\ and\ \bibinfo {author} {\bibfnamefont {C.~C.}\ \bibnamefont
  {Rice}},\ }\bibfield  {title} {\bibinfo {title} {Minimal model
  explanations},\ }\href@noop {} {\bibfield  {journal} {\bibinfo  {journal}
  {Philosophy of Science}\ }\textbf {\bibinfo {volume} {81}},\ \bibinfo {pages}
  {349} (\bibinfo {year} {2014})}\BibitemShut {NoStop}%
\bibitem [{\citenamefont {Oono}(2012)}]{oono2012nonlinear}%
  \BibitemOpen
  \bibfield  {author} {\bibinfo {author} {\bibfnamefont {Y.}~\bibnamefont
  {Oono}},\ }\href@noop {} {\emph {\bibinfo {title} {The nonlinear world:
  Conceptual analysis and phenomenology}}}\ (\bibinfo  {publisher} {Springer
  Science \& Business Media},\ \bibinfo {year} {2012})\BibitemShut {NoStop}%
\bibitem [{\citenamefont {Hoel}\ \emph {et~al.}(2013)\citenamefont {Hoel},
  \citenamefont {Albantakis},\ and\ \citenamefont
  {Tononi}}]{hoel2013quantifying}%
  \BibitemOpen
  \bibfield  {author} {\bibinfo {author} {\bibfnamefont {E.~P.}\ \bibnamefont
  {Hoel}}, \bibinfo {author} {\bibfnamefont {L.}~\bibnamefont {Albantakis}},\
  and\ \bibinfo {author} {\bibfnamefont {G.}~\bibnamefont {Tononi}},\
  }\bibfield  {title} {\bibinfo {title} {Quantifying causal emergence shows
  that macro can beat micro},\ }\href@noop {} {\bibfield  {journal} {\bibinfo
  {journal} {Proceedings of the National Academy of Sciences}\ }\textbf
  {\bibinfo {volume} {110}},\ \bibinfo {pages} {19790} (\bibinfo {year}
  {2013})}\BibitemShut {NoStop}%
\bibitem [{\citenamefont {Machta}\ \emph {et~al.}(2013)\citenamefont {Machta},
  \citenamefont {Chachra}, \citenamefont {Transtrum},\ and\ \citenamefont
  {Sethna}}]{machta2013parameter}%
  \BibitemOpen
  \bibfield  {author} {\bibinfo {author} {\bibfnamefont {B.~B.}\ \bibnamefont
  {Machta}}, \bibinfo {author} {\bibfnamefont {R.}~\bibnamefont {Chachra}},
  \bibinfo {author} {\bibfnamefont {M.~K.}\ \bibnamefont {Transtrum}},\ and\
  \bibinfo {author} {\bibfnamefont {J.~P.}\ \bibnamefont {Sethna}},\ }\bibfield
   {title} {\bibinfo {title} {Parameter space compression underlies emergent
  theories and predictive models},\ }\href@noop {} {\bibfield  {journal}
  {\bibinfo  {journal} {Science}\ }\textbf {\bibinfo {volume} {342}},\ \bibinfo
  {pages} {604} (\bibinfo {year} {2013})}\BibitemShut {NoStop}%
\bibitem [{\citenamefont {Mattingly}\ \emph {et~al.}(2018)\citenamefont
  {Mattingly}, \citenamefont {Transtrum}, \citenamefont {Abbott},\ and\
  \citenamefont {Machta}}]{mattingly2018maximizing}%
  \BibitemOpen
  \bibfield  {author} {\bibinfo {author} {\bibfnamefont {H.~H.}\ \bibnamefont
  {Mattingly}}, \bibinfo {author} {\bibfnamefont {M.~K.}\ \bibnamefont
  {Transtrum}}, \bibinfo {author} {\bibfnamefont {M.~C.}\ \bibnamefont
  {Abbott}},\ and\ \bibinfo {author} {\bibfnamefont {B.~B.}\ \bibnamefont
  {Machta}},\ }\bibfield  {title} {\bibinfo {title} {Maximizing the information
  learned from finite data selects a simple model},\ }\href@noop {} {\bibfield
  {journal} {\bibinfo  {journal} {Proceedings of the National Academy of
  Sciences}\ }\textbf {\bibinfo {volume} {115}},\ \bibinfo {pages} {1760}
  (\bibinfo {year} {2018})}\BibitemShut {NoStop}%
\bibitem [{\citenamefont {Gordon}\ \emph {et~al.}(2021)\citenamefont {Gordon},
  \citenamefont {Banerjee}, \citenamefont {Koch-Janusz},\ and\ \citenamefont
  {Ringel}}]{gordon2021relevance}%
  \BibitemOpen
  \bibfield  {author} {\bibinfo {author} {\bibfnamefont {A.}~\bibnamefont
  {Gordon}}, \bibinfo {author} {\bibfnamefont {A.}~\bibnamefont {Banerjee}},
  \bibinfo {author} {\bibfnamefont {M.}~\bibnamefont {Koch-Janusz}},\ and\
  \bibinfo {author} {\bibfnamefont {Z.}~\bibnamefont {Ringel}},\ }\bibfield
  {title} {\bibinfo {title} {Relevance in the renormalization group and in
  information theory},\ }\href@noop {} {\bibfield  {journal} {\bibinfo
  {journal} {Physical Review Letters}\ }\textbf {\bibinfo {volume} {126}},\
  \bibinfo {pages} {240601} (\bibinfo {year} {2021})}\BibitemShut {NoStop}%
\bibitem [{\citenamefont {Akaike}(1974)}]{akaike1974new}%
  \BibitemOpen
  \bibfield  {author} {\bibinfo {author} {\bibfnamefont {H.}~\bibnamefont
  {Akaike}},\ }\bibfield  {title} {\bibinfo {title} {A new look at the
  statistical model identification},\ }\href@noop {} {\bibfield  {journal}
  {\bibinfo  {journal} {IEEE transactions on automatic control}\ }\textbf
  {\bibinfo {volume} {19}},\ \bibinfo {pages} {716} (\bibinfo {year}
  {1974})}\BibitemShut {NoStop}%
\bibitem [{\citenamefont {Schwarz}\ \emph {et~al.}(1978)\citenamefont {Schwarz}
  \emph {et~al.}}]{schwarz1978estimating}%
  \BibitemOpen
  \bibfield  {author} {\bibinfo {author} {\bibfnamefont {G.}~\bibnamefont
  {Schwarz}} \emph {et~al.},\ }\bibfield  {title} {\bibinfo {title} {Estimating
  the dimension of a model},\ }\href@noop {} {\bibfield  {journal} {\bibinfo
  {journal} {Annals of statistics}\ }\textbf {\bibinfo {volume} {6}},\ \bibinfo
  {pages} {461} (\bibinfo {year} {1978})}\BibitemShut {NoStop}%
\bibitem [{\citenamefont {Trotta}(2008)}]{trotta2008bayes}%
  \BibitemOpen
  \bibfield  {author} {\bibinfo {author} {\bibfnamefont {R.}~\bibnamefont
  {Trotta}},\ }\bibfield  {title} {\bibinfo {title} {Bayes in the sky: Bayesian
  inference and model selection in cosmology},\ }\href@noop {} {\bibfield
  {journal} {\bibinfo  {journal} {Contemporary Physics}\ }\textbf {\bibinfo
  {volume} {49}},\ \bibinfo {pages} {71} (\bibinfo {year} {2008})}\BibitemShut
  {NoStop}%
\bibitem [{\citenamefont {Mann}(2011)}]{mann2011bayesian}%
  \BibitemOpen
  \bibfield  {author} {\bibinfo {author} {\bibfnamefont {R.~P.}\ \bibnamefont
  {Mann}},\ }\bibfield  {title} {\bibinfo {title} {Bayesian inference for
  identifying interaction rules in moving animal groups},\ }\href@noop {}
  {\bibfield  {journal} {\bibinfo  {journal} {PloS one}\ }\textbf {\bibinfo
  {volume} {6}},\ \bibinfo {pages} {e22827} (\bibinfo {year}
  {2011})}\BibitemShut {NoStop}%
\bibitem [{\citenamefont {Mark}\ \emph {et~al.}(2018)\citenamefont {Mark},
  \citenamefont {Metzner}, \citenamefont {Lautscham}, \citenamefont {Strissel},
  \citenamefont {Strick},\ and\ \citenamefont {Fabry}}]{mark2018bayesian}%
  \BibitemOpen
  \bibfield  {author} {\bibinfo {author} {\bibfnamefont {C.}~\bibnamefont
  {Mark}}, \bibinfo {author} {\bibfnamefont {C.}~\bibnamefont {Metzner}},
  \bibinfo {author} {\bibfnamefont {L.}~\bibnamefont {Lautscham}}, \bibinfo
  {author} {\bibfnamefont {P.~L.}\ \bibnamefont {Strissel}}, \bibinfo {author}
  {\bibfnamefont {R.}~\bibnamefont {Strick}},\ and\ \bibinfo {author}
  {\bibfnamefont {B.}~\bibnamefont {Fabry}},\ }\bibfield  {title} {\bibinfo
  {title} {Bayesian model selection for complex dynamic systems},\ }\href@noop
  {} {\bibfield  {journal} {\bibinfo  {journal} {Nature communications}\
  }\textbf {\bibinfo {volume} {9}},\ \bibinfo {pages} {1} (\bibinfo {year}
  {2018})}\BibitemShut {NoStop}%
\bibitem [{\citenamefont {V{\'a}zquez}\ \emph {et~al.}(2021)\citenamefont
  {V{\'a}zquez}, \citenamefont {Tamayo}, \citenamefont {Sen},\ and\
  \citenamefont {Quiros}}]{vazquez2021bayesian}%
  \BibitemOpen
  \bibfield  {author} {\bibinfo {author} {\bibfnamefont {J.~A.}\ \bibnamefont
  {V{\'a}zquez}}, \bibinfo {author} {\bibfnamefont {D.}~\bibnamefont {Tamayo}},
  \bibinfo {author} {\bibfnamefont {A.~A.}\ \bibnamefont {Sen}},\ and\ \bibinfo
  {author} {\bibfnamefont {I.}~\bibnamefont {Quiros}},\ }\bibfield  {title}
  {\bibinfo {title} {Bayesian model selection on scalar $\varepsilon$-field
  dark energy},\ }\href@noop {} {\bibfield  {journal} {\bibinfo  {journal}
  {Physical Review D}\ }\textbf {\bibinfo {volume} {103}},\ \bibinfo {pages}
  {043506} (\bibinfo {year} {2021})}\BibitemShut {NoStop}%
\bibitem [{\citenamefont {Tokuda}\ \emph
  {et~al.}(2021{\natexlab{a}})\citenamefont {Tokuda}, \citenamefont {Souma},
  \citenamefont {Segawa}, \citenamefont {Takahashi}, \citenamefont {Ando},
  \citenamefont {Nakanishi},\ and\ \citenamefont {Sato}}]{tokuda2021unveiling}%
  \BibitemOpen
  \bibfield  {author} {\bibinfo {author} {\bibfnamefont {S.}~\bibnamefont
  {Tokuda}}, \bibinfo {author} {\bibfnamefont {S.}~\bibnamefont {Souma}},
  \bibinfo {author} {\bibfnamefont {K.}~\bibnamefont {Segawa}}, \bibinfo
  {author} {\bibfnamefont {T.}~\bibnamefont {Takahashi}}, \bibinfo {author}
  {\bibfnamefont {Y.}~\bibnamefont {Ando}}, \bibinfo {author} {\bibfnamefont
  {T.}~\bibnamefont {Nakanishi}},\ and\ \bibinfo {author} {\bibfnamefont
  {T.}~\bibnamefont {Sato}},\ }\bibfield  {title} {\bibinfo {title} {Unveiling
  quasiparticle dynamics of topological insulators through bayesian
  modelling},\ }\href@noop {} {\bibfield  {journal} {\bibinfo  {journal}
  {Communications Physics}\ }\textbf {\bibinfo {volume} {4}},\ \bibinfo {pages}
  {1} (\bibinfo {year} {2021}{\natexlab{a}})}\BibitemShut {NoStop}%
\bibitem [{\citenamefont {Tokuda}\ \emph
  {et~al.}(2021{\natexlab{b}})\citenamefont {Tokuda}, \citenamefont {Kawachi},
  \citenamefont {Sasaki}, \citenamefont {Arakawa}, \citenamefont {Yamasaki},
  \citenamefont {Terasaka},\ and\ \citenamefont
  {Inagaki}}]{tokuda2021bayesian}%
  \BibitemOpen
  \bibfield  {author} {\bibinfo {author} {\bibfnamefont {S.}~\bibnamefont
  {Tokuda}}, \bibinfo {author} {\bibfnamefont {Y.}~\bibnamefont {Kawachi}},
  \bibinfo {author} {\bibfnamefont {M.}~\bibnamefont {Sasaki}}, \bibinfo
  {author} {\bibfnamefont {H.}~\bibnamefont {Arakawa}}, \bibinfo {author}
  {\bibfnamefont {K.}~\bibnamefont {Yamasaki}}, \bibinfo {author}
  {\bibfnamefont {K.}~\bibnamefont {Terasaka}},\ and\ \bibinfo {author}
  {\bibfnamefont {S.}~\bibnamefont {Inagaki}},\ }\bibfield  {title} {\bibinfo
  {title} {Bayesian inference of ion velocity distribution function from
  laser-induced fluorescence spectra},\ }\href@noop {} {\bibfield  {journal}
  {\bibinfo  {journal} {Scientific Reports}\ }\textbf {\bibinfo {volume}
  {11}},\ \bibinfo {pages} {1} (\bibinfo {year}
  {2021}{\natexlab{b}})}\BibitemShut {NoStop}%
\bibitem [{\citenamefont {Wasserman}(2000)}]{wasserman2000bayesian}%
  \BibitemOpen
  \bibfield  {author} {\bibinfo {author} {\bibfnamefont {L.}~\bibnamefont
  {Wasserman}},\ }\bibfield  {title} {\bibinfo {title} {Bayesian model
  selection and model averaging},\ }\href@noop {} {\bibfield  {journal}
  {\bibinfo  {journal} {Journal of mathematical psychology}\ }\textbf {\bibinfo
  {volume} {44}},\ \bibinfo {pages} {92} (\bibinfo {year} {2000})}\BibitemShut
  {NoStop}%
\bibitem [{\citenamefont {Berger}\ \emph {et~al.}(2001)\citenamefont {Berger},
  \citenamefont {Pericchi}, \citenamefont {Ghosh}, \citenamefont {Samanta},
  \citenamefont {De~Santis}, \citenamefont {Berger},\ and\ \citenamefont
  {Pericchi}}]{berger2001objective}%
  \BibitemOpen
  \bibfield  {author} {\bibinfo {author} {\bibfnamefont {J.~O.}\ \bibnamefont
  {Berger}}, \bibinfo {author} {\bibfnamefont {L.~R.}\ \bibnamefont
  {Pericchi}}, \bibinfo {author} {\bibfnamefont {J.}~\bibnamefont {Ghosh}},
  \bibinfo {author} {\bibfnamefont {T.}~\bibnamefont {Samanta}}, \bibinfo
  {author} {\bibfnamefont {F.}~\bibnamefont {De~Santis}}, \bibinfo {author}
  {\bibfnamefont {J.}~\bibnamefont {Berger}},\ and\ \bibinfo {author}
  {\bibfnamefont {L.}~\bibnamefont {Pericchi}},\ }\bibfield  {title} {\bibinfo
  {title} {Objective bayesian methods for model selection: Introduction and
  comparison},\ }\href@noop {} {\bibfield  {journal} {\bibinfo  {journal}
  {Lecture Notes-Monograph Series}\ ,\ \bibinfo {pages} {135}} (\bibinfo {year}
  {2001})}\BibitemShut {NoStop}%
\bibitem [{\citenamefont {MacKay}(1992)}]{mackay1992bayesian}%
  \BibitemOpen
  \bibfield  {author} {\bibinfo {author} {\bibfnamefont {D.~J.}\ \bibnamefont
  {MacKay}},\ }\bibfield  {title} {\bibinfo {title} {Bayesian interpolation},\
  }\href@noop {} {\bibfield  {journal} {\bibinfo  {journal} {Neural
  computation}\ }\textbf {\bibinfo {volume} {4}},\ \bibinfo {pages} {415}
  (\bibinfo {year} {1992})}\BibitemShut {NoStop}%
\bibitem [{\citenamefont
  {Balasubramanian}(1997)}]{balasubramanian1997statistical}%
  \BibitemOpen
  \bibfield  {author} {\bibinfo {author} {\bibfnamefont {V.}~\bibnamefont
  {Balasubramanian}},\ }\bibfield  {title} {\bibinfo {title} {Statistical
  inference, occam's razor, and statistical mechanics on the space of
  probability distributions},\ }\href@noop {} {\bibfield  {journal} {\bibinfo
  {journal} {Neural computation}\ }\textbf {\bibinfo {volume} {9}},\ \bibinfo
  {pages} {349} (\bibinfo {year} {1997})}\BibitemShut {NoStop}%
\bibitem [{\citenamefont {Watanabe}(2001)}]{watanabe2001algebraic}%
  \BibitemOpen
  \bibfield  {author} {\bibinfo {author} {\bibfnamefont {S.}~\bibnamefont
  {Watanabe}},\ }\bibfield  {title} {\bibinfo {title} {Algebraic analysis for
  nonidentifiable learning machines},\ }\href@noop {} {\bibfield  {journal}
  {\bibinfo  {journal} {Neural Computation}\ }\textbf {\bibinfo {volume}
  {13}},\ \bibinfo {pages} {899} (\bibinfo {year} {2001})}\BibitemShut
  {NoStop}%
\bibitem [{\citenamefont {Watanabe}(2009)}]{watanabe2009algebraic}%
  \BibitemOpen
  \bibfield  {author} {\bibinfo {author} {\bibfnamefont {S.}~\bibnamefont
  {Watanabe}},\ }\href@noop {} {\emph {\bibinfo {title} {Algebraic geometry and
  statistical learning theory}}},\ Vol.~\bibinfo {volume} {25}\ (\bibinfo
  {publisher} {Cambridge University Press},\ \bibinfo {year}
  {2009})\BibitemShut {NoStop}%
\bibitem [{\citenamefont {Watanabe}(2018)}]{watanabe2018mathematical}%
  \BibitemOpen
  \bibfield  {author} {\bibinfo {author} {\bibfnamefont {S.}~\bibnamefont
  {Watanabe}},\ }\href@noop {} {\emph {\bibinfo {title} {Mathematical theory of
  Bayesian statistics}}}\ (\bibinfo  {publisher} {CRC Press},\ \bibinfo {year}
  {2018})\BibitemShut {NoStop}%
\bibitem [{\citenamefont {Watanabe}(2013)}]{watanabe2013widely}%
  \BibitemOpen
  \bibfield  {author} {\bibinfo {author} {\bibfnamefont {S.}~\bibnamefont
  {Watanabe}},\ }\bibfield  {title} {\bibinfo {title} {A widely applicable
  bayesian information criterion},\ }\href@noop {} {\bibfield  {journal}
  {\bibinfo  {journal} {Journal of Machine Learning Research}\ }\textbf
  {\bibinfo {volume} {14}},\ \bibinfo {pages} {867} (\bibinfo {year}
  {2013})}\BibitemShut {NoStop}%
\bibitem [{\citenamefont {Tokuda}\ \emph {et~al.}(2014)\citenamefont {Tokuda},
  \citenamefont {Nagata},\ and\ \citenamefont {Okada}}]{tokuda2014numerical}%
  \BibitemOpen
  \bibfield  {author} {\bibinfo {author} {\bibfnamefont {S.}~\bibnamefont
  {Tokuda}}, \bibinfo {author} {\bibfnamefont {K.}~\bibnamefont {Nagata}},\
  and\ \bibinfo {author} {\bibfnamefont {M.}~\bibnamefont {Okada}},\ }\bibfield
   {title} {\bibinfo {title} {A numerical analysis of learning coefficient in
  radial basis function network},\ }\href@noop {} {\bibfield  {journal}
  {\bibinfo  {journal} {IPSJ Online Transactions}\ }\textbf {\bibinfo {volume}
  {7}},\ \bibinfo {pages} {20} (\bibinfo {year} {2014})}\BibitemShut {NoStop}%
\bibitem [{\citenamefont {Jaynes}(1957)}]{jaynes1957information}%
  \BibitemOpen
  \bibfield  {author} {\bibinfo {author} {\bibfnamefont {E.~T.}\ \bibnamefont
  {Jaynes}},\ }\bibfield  {title} {\bibinfo {title} {Information theory and
  statistical mechanics},\ }\href@noop {} {\bibfield  {journal} {\bibinfo
  {journal} {Physical review}\ }\textbf {\bibinfo {volume} {106}},\ \bibinfo
  {pages} {620} (\bibinfo {year} {1957})}\BibitemShut {NoStop}%
\bibitem [{\citenamefont {Jaynes}(2003)}]{jaynes2003probability}%
  \BibitemOpen
  \bibfield  {author} {\bibinfo {author} {\bibfnamefont {E.~T.}\ \bibnamefont
  {Jaynes}},\ }\href@noop {} {\emph {\bibinfo {title} {Probability theory: The
  logic of science}}}\ (\bibinfo  {publisher} {Cambridge university press},\
  \bibinfo {year} {2003})\BibitemShut {NoStop}%
\bibitem [{\citenamefont {Zdeborov{\'a}}\ and\ \citenamefont
  {Krzakala}(2016)}]{zdeborova2016statistical}%
  \BibitemOpen
  \bibfield  {author} {\bibinfo {author} {\bibfnamefont {L.}~\bibnamefont
  {Zdeborov{\'a}}}\ and\ \bibinfo {author} {\bibfnamefont {F.}~\bibnamefont
  {Krzakala}},\ }\bibfield  {title} {\bibinfo {title} {Statistical physics of
  inference: Thresholds and algorithms},\ }\href@noop {} {\bibfield  {journal}
  {\bibinfo  {journal} {Advances in Physics}\ }\textbf {\bibinfo {volume}
  {65}},\ \bibinfo {pages} {453} (\bibinfo {year} {2016})}\BibitemShut
  {NoStop}%
\bibitem [{\citenamefont {Efron}\ and\ \citenamefont
  {Morris}(1973)}]{efron1973stein}%
  \BibitemOpen
  \bibfield  {author} {\bibinfo {author} {\bibfnamefont {B.}~\bibnamefont
  {Efron}}\ and\ \bibinfo {author} {\bibfnamefont {C.}~\bibnamefont {Morris}},\
  }\bibfield  {title} {\bibinfo {title} {Stein's estimation rule and its
  competitors―an empirical bayes approach},\ }\href@noop {} {\bibfield
  {journal} {\bibinfo  {journal} {Journal of the American Statistical
  Association}\ }\textbf {\bibinfo {volume} {68}},\ \bibinfo {pages} {117}
  (\bibinfo {year} {1973})}\BibitemShut {NoStop}%
\bibitem [{\citenamefont {Akaike}(1998)}]{akaike1998likelihood}%
  \BibitemOpen
  \bibfield  {author} {\bibinfo {author} {\bibfnamefont {H.}~\bibnamefont
  {Akaike}},\ }\bibfield  {title} {\bibinfo {title} {Likelihood and the bayes
  procedure},\ }in\ \href@noop {} {\emph {\bibinfo {booktitle} {Selected Papers
  of Hirotugu Akaike}}}\ (\bibinfo  {publisher} {Springer},\ \bibinfo {year}
  {1998})\ pp.\ \bibinfo {pages} {309--332}\BibitemShut {NoStop}%
\bibitem [{\citenamefont {Tokuda}\ \emph {et~al.}(2017)\citenamefont {Tokuda},
  \citenamefont {Nagata},\ and\ \citenamefont
  {Okada}}]{tokuda2017simultaneous}%
  \BibitemOpen
  \bibfield  {author} {\bibinfo {author} {\bibfnamefont {S.}~\bibnamefont
  {Tokuda}}, \bibinfo {author} {\bibfnamefont {K.}~\bibnamefont {Nagata}},\
  and\ \bibinfo {author} {\bibfnamefont {M.}~\bibnamefont {Okada}},\ }\bibfield
   {title} {\bibinfo {title} {Simultaneous estimation of noise variance and
  number of peaks in bayesian spectral deconvolution},\ }\href@noop {}
  {\bibfield  {journal} {\bibinfo  {journal} {Journal of the Physical Society
  of Japan}\ }\textbf {\bibinfo {volume} {86}},\ \bibinfo {pages} {024001}
  (\bibinfo {year} {2017})}\BibitemShut {NoStop}%
\bibitem [{\citenamefont {Nishimori}(1980)}]{nishimori1980exact}%
  \BibitemOpen
  \bibfield  {author} {\bibinfo {author} {\bibfnamefont {H.}~\bibnamefont
  {Nishimori}},\ }\bibfield  {title} {\bibinfo {title} {Exact results and
  critical properties of the ising model with competing interactions},\
  }\href@noop {} {\bibfield  {journal} {\bibinfo  {journal} {Journal of Physics
  C: Solid State Physics}\ }\textbf {\bibinfo {volume} {13}},\ \bibinfo {pages}
  {4071} (\bibinfo {year} {1980})}\BibitemShut {NoStop}%
\bibitem [{\citenamefont {Iba}(1999)}]{iba1999nishimori}%
  \BibitemOpen
  \bibfield  {author} {\bibinfo {author} {\bibfnamefont {Y.}~\bibnamefont
  {Iba}},\ }\bibfield  {title} {\bibinfo {title} {The nishimori line and
  bayesian statistics},\ }\href@noop {} {\bibfield  {journal} {\bibinfo
  {journal} {Journal of Physics A: Mathematical and General}\ }\textbf
  {\bibinfo {volume} {32}},\ \bibinfo {pages} {3875} (\bibinfo {year}
  {1999})}\BibitemShut {NoStop}%
\bibitem [{\citenamefont
  {Watanabe}(2010{\natexlab{a}})}]{watanabe2010equations}%
  \BibitemOpen
  \bibfield  {author} {\bibinfo {author} {\bibfnamefont {S.}~\bibnamefont
  {Watanabe}},\ }\bibfield  {title} {\bibinfo {title} {Equations of states in
  singular statistical estimation},\ }\href@noop {} {\bibfield  {journal}
  {\bibinfo  {journal} {Neural Networks}\ }\textbf {\bibinfo {volume} {23}},\
  \bibinfo {pages} {20} (\bibinfo {year} {2010}{\natexlab{a}})}\BibitemShut
  {NoStop}%
\bibitem [{\citenamefont
  {Watanabe}(2010{\natexlab{b}})}]{watanabe2010asymptotic}%
  \BibitemOpen
  \bibfield  {author} {\bibinfo {author} {\bibfnamefont {S.}~\bibnamefont
  {Watanabe}},\ }\bibfield  {title} {\bibinfo {title} {Asymptotic equivalence
  of bayes cross validation and widely applicable information criterion in
  singular learning theory},\ }\href@noop {} {\bibfield  {journal} {\bibinfo
  {journal} {Journal of Machine Learning Research}\ }\textbf {\bibinfo {volume}
  {11}},\ \bibinfo {pages} {3571} (\bibinfo {year}
  {2010}{\natexlab{b}})}\BibitemShut {NoStop}%
\bibitem [{\citenamefont {Watanabe}\ \emph {et~al.}(2010)\citenamefont
  {Watanabe} \emph {et~al.}}]{watanabe2010limit}%
  \BibitemOpen
  \bibfield  {author} {\bibinfo {author} {\bibfnamefont {S.}~\bibnamefont
  {Watanabe}} \emph {et~al.},\ }\bibfield  {title} {\bibinfo {title} {A limit
  theorem in singular regression problem},\ }in\ \href@noop {} {\emph {\bibinfo
  {booktitle} {Probabilistic Approach to Geometry}}}\ (\bibinfo {organization}
  {Mathematical Society of Japan},\ \bibinfo {year} {2010})\ pp.\ \bibinfo
  {pages} {473--492}\BibitemShut {NoStop}%
\bibitem [{\citenamefont {BROOMHEAD}(1988)}]{broomhead1988multivariable}%
  \BibitemOpen
  \bibfield  {author} {\bibinfo {author} {\bibfnamefont {D.}~\bibnamefont
  {BROOMHEAD}},\ }\bibfield  {title} {\bibinfo {title} {Multivariable
  functional interpolation and adaptive networks},\ }\href@noop {} {\bibfield
  {journal} {\bibinfo  {journal} {Complex Systems}\ }\textbf {\bibinfo {volume}
  {2}},\ \bibinfo {pages} {321} (\bibinfo {year} {1988})}\BibitemShut {NoStop}%
\bibitem [{\citenamefont {Geyer}(1991)}]{geyer1991markov}%
  \BibitemOpen
  \bibfield  {author} {\bibinfo {author} {\bibfnamefont {C.~J.}\ \bibnamefont
  {Geyer}},\ }\bibfield  {title} {\bibinfo {title} {Markov chain monte carlo
  maximum likelihood}\ }(\bibinfo  {publisher} {Interface Foundation of North
  America},\ \bibinfo {year} {1991})\BibitemShut {NoStop}%
\bibitem [{\citenamefont {Hukushima}\ and\ \citenamefont
  {Nemoto}(1996)}]{hukushima1996exchange}%
  \BibitemOpen
  \bibfield  {author} {\bibinfo {author} {\bibfnamefont {K.}~\bibnamefont
  {Hukushima}}\ and\ \bibinfo {author} {\bibfnamefont {K.}~\bibnamefont
  {Nemoto}},\ }\bibfield  {title} {\bibinfo {title} {Exchange monte carlo
  method and application to spin glass simulations},\ }\href@noop {} {\bibfield
   {journal} {\bibinfo  {journal} {Journal of the Physical Society of Japan}\
  }\textbf {\bibinfo {volume} {65}},\ \bibinfo {pages} {1604} (\bibinfo {year}
  {1996})}\BibitemShut {NoStop}%
\bibitem [{\citenamefont {Meng}\ and\ \citenamefont
  {Wong}(1996)}]{meng1996simulating}%
  \BibitemOpen
  \bibfield  {author} {\bibinfo {author} {\bibfnamefont {X.-L.}\ \bibnamefont
  {Meng}}\ and\ \bibinfo {author} {\bibfnamefont {W.~H.}\ \bibnamefont
  {Wong}},\ }\bibfield  {title} {\bibinfo {title} {Simulating ratios of
  normalizing constants via a simple identity: a theoretical exploration},\
  }\href@noop {} {\bibfield  {journal} {\bibinfo  {journal} {Statistica
  Sinica}\ ,\ \bibinfo {pages} {831}} (\bibinfo {year} {1996})}\BibitemShut
  {NoStop}%
\bibitem [{\citenamefont {Gelman}\ and\ \citenamefont
  {Meng}(1998)}]{gelman1998simulating}%
  \BibitemOpen
  \bibfield  {author} {\bibinfo {author} {\bibfnamefont {A.}~\bibnamefont
  {Gelman}}\ and\ \bibinfo {author} {\bibfnamefont {X.-L.}\ \bibnamefont
  {Meng}},\ }\bibfield  {title} {\bibinfo {title} {Simulating normalizing
  constants: From importance sampling to bridge sampling to path sampling},\
  }\href@noop {} {\bibfield  {journal} {\bibinfo  {journal} {Statistical
  science}\ ,\ \bibinfo {pages} {163}} (\bibinfo {year} {1998})}\BibitemShut
  {NoStop}%
\bibitem [{\citenamefont {LaMont}\ and\ \citenamefont
  {Wiggins}(2019)}]{lamont2019correspondence}%
  \BibitemOpen
  \bibfield  {author} {\bibinfo {author} {\bibfnamefont {C.~H.}\ \bibnamefont
  {LaMont}}\ and\ \bibinfo {author} {\bibfnamefont {P.~A.}\ \bibnamefont
  {Wiggins}},\ }\bibfield  {title} {\bibinfo {title} {Correspondence between
  thermodynamics and inference},\ }\href@noop {} {\bibfield  {journal}
  {\bibinfo  {journal} {Physical Review E}\ }\textbf {\bibinfo {volume} {99}},\
  \bibinfo {pages} {052140} (\bibinfo {year} {2019})}\BibitemShut {NoStop}%
\end{thebibliography}
%

\newpage

\newpage

\newpage

\newpage

\newpage

\newpage

\renewcommand{\thefigure}{A\arabic{figure} }
\setcounter{figure}{0}

\begin{figure*}[htbp]
\begin{center}
\begin{tabular}{ll}
{\LARGE {\bf a}} & {\LARGE {\bf b}} \\
\includegraphics[width=8.6cm]{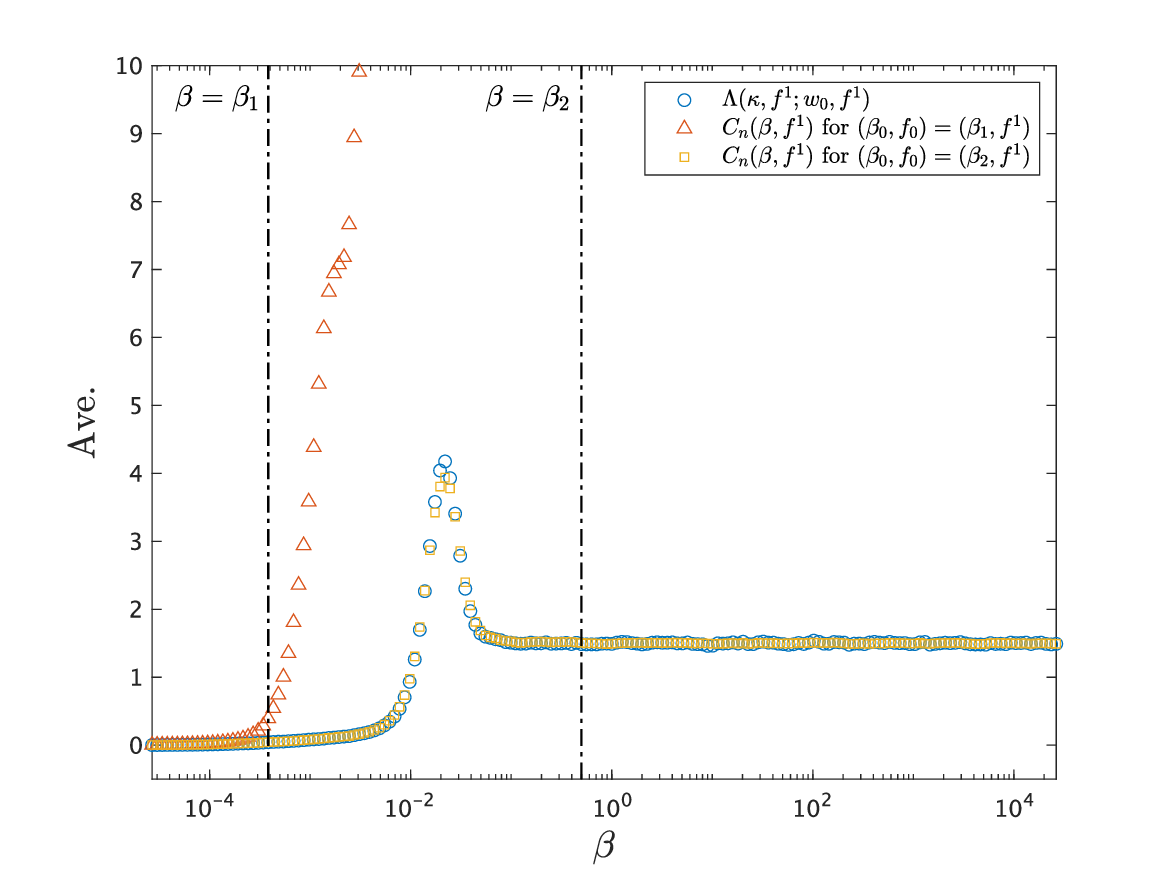}
&
\includegraphics[width=8.6cm]{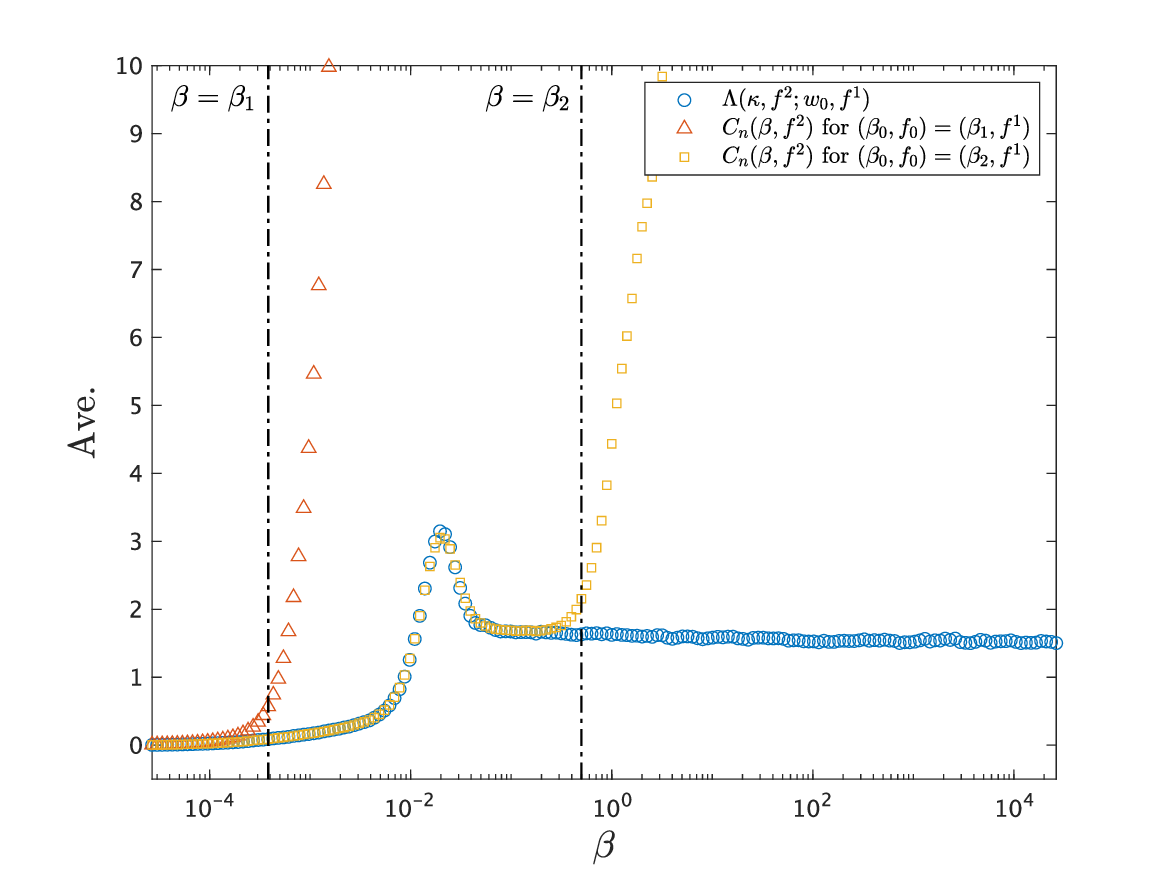} \\
{\LARGE {\bf c}} & {\LARGE {\bf d}} \\
\includegraphics[width=8.6cm]{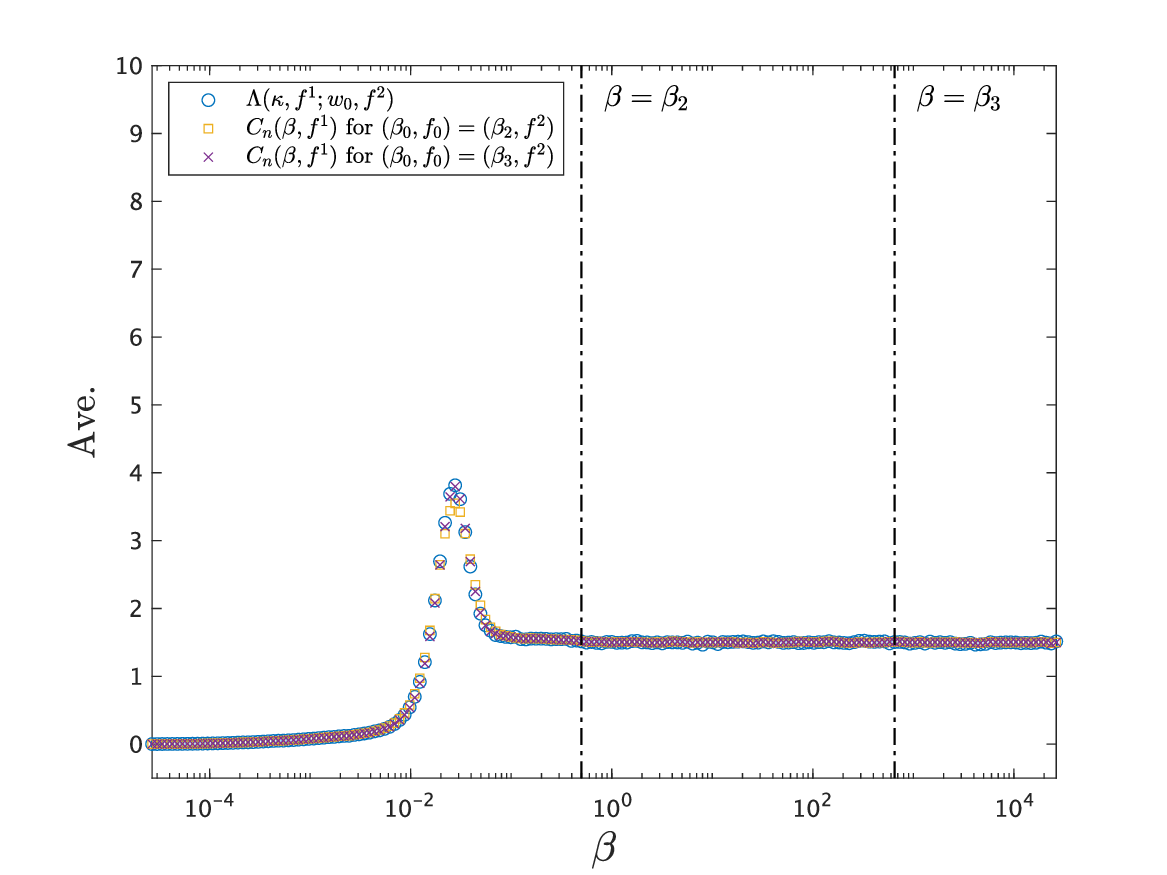}
&
\includegraphics[width=8.6cm]{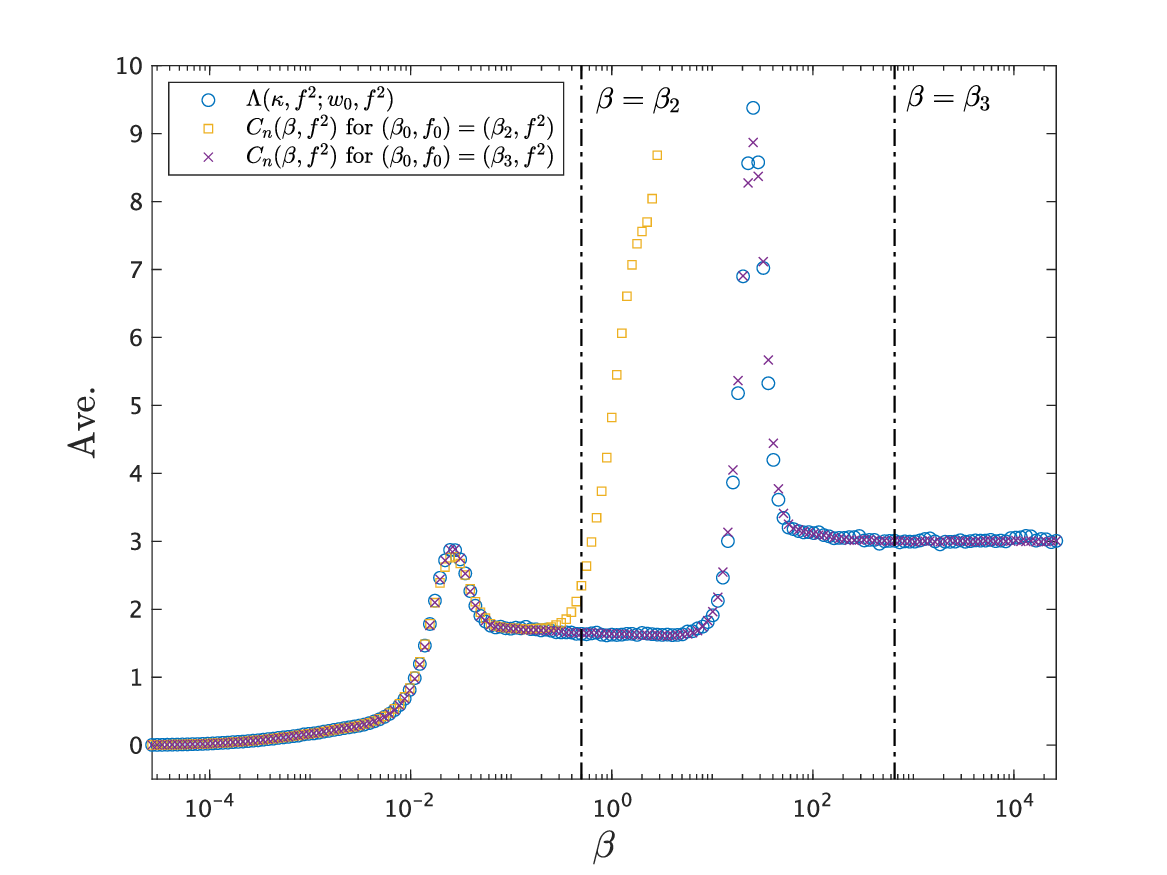}
\end{tabular}
\end{center}
\caption{Finite-size effects on the Bayes specific heat. Semi-log plots of $\Lambda(\kappa, f; w_0, f_0)$ (blue circles) and the expectations of $C_n(\beta,f)$ over realizations of $D^n$ for $\beta_0=\beta_1$ (red triangles), $\beta_0=\beta_2$ (yellow squares) and $\beta_0=\beta_3$ (purple crosses), where $\beta=\kappa/n$ ($\beta_1:=\kappa_1/n$, $\beta_2:=\kappa_2 /n$, and $\beta_3:=\kappa_3 /n$).
For the case that the ground truth is $f_0=f^1$ with $w_0 = \{10, 10, 0\}$, ({\bf a}) $C_n(\beta,f^1)$ and ({\bf b}) $C_n(\beta,f^2)$ are respectively compared with $\Lambda(\kappa, f^1; w_0, f^1)$ and $\Lambda(\kappa, f^2; w_0, f^1)$ shown in Fig. 2b.
For the case that the ground truth is $f_0=f^1$ with $w_0 = \{5, 10, (-1)^k \times 0.25 \}_{k=1}^2$, ({\bf c}) $C_n(\beta,f^1)$ and ({\bf d}) $C_n(\beta,f^2)$ are respectively compared with $\Lambda(\kappa, f^1; w_0, f^2)$ and $\Lambda(\kappa, f^2; w_0, f^2)$ shown in Fig. 2d.}
\label{fig:Fig.S1}
\end{figure*}

\newpage

\begin{figure}[htbp]
\begin{center}
\begin{tabular}{c}
\includegraphics[width=8.6cm]{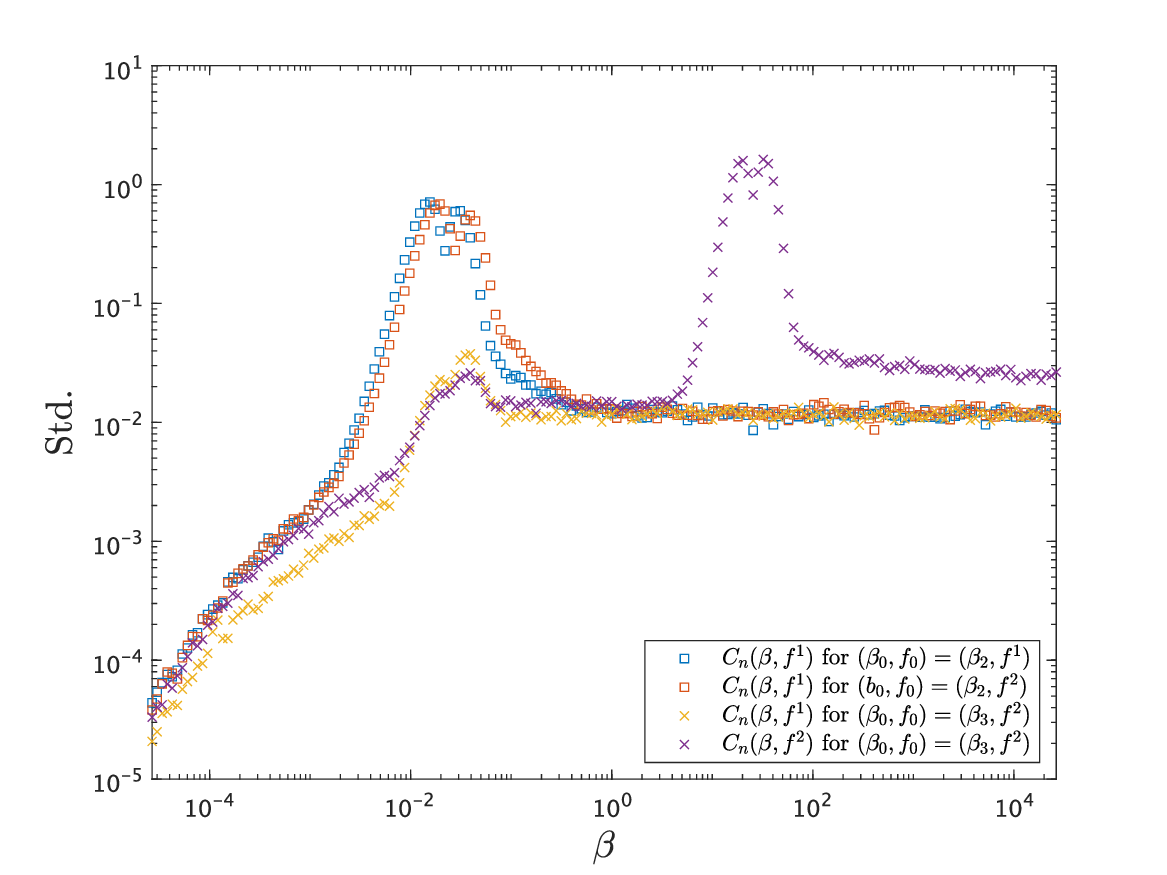}
\end{tabular}
\end{center}
\caption{Fluctuation of the Bayes specific heat for realizations. Log-log plots of the standard deviation of $C_n(\beta,f)$ for realizations of $D^n$ for $(\beta_0,f_0)=(\beta_2,f^1)$ (blue squares for $f=f^1$), $(\beta_0,f_0)=(\beta_2, f^2)$ (red squares for $f=f^1$), and $(\beta_0,f_0)=(\beta_3, f^2)$ (yellow crosses for $f=f^1$ and purple crosses for $f=f^2$), where each case corresponds to Fig. \ref{fig:Fig.S1}.}
\label{fig:Fig.S2}
\end{figure}

\newpage

\begin{figure*}[htbp]
\begin{center}
\begin{tabular}{lll}
{\LARGE {\bf a}} & {\LARGE {\bf b}} & {\LARGE {\bf c}} \\
\includegraphics[width=5.7cm]{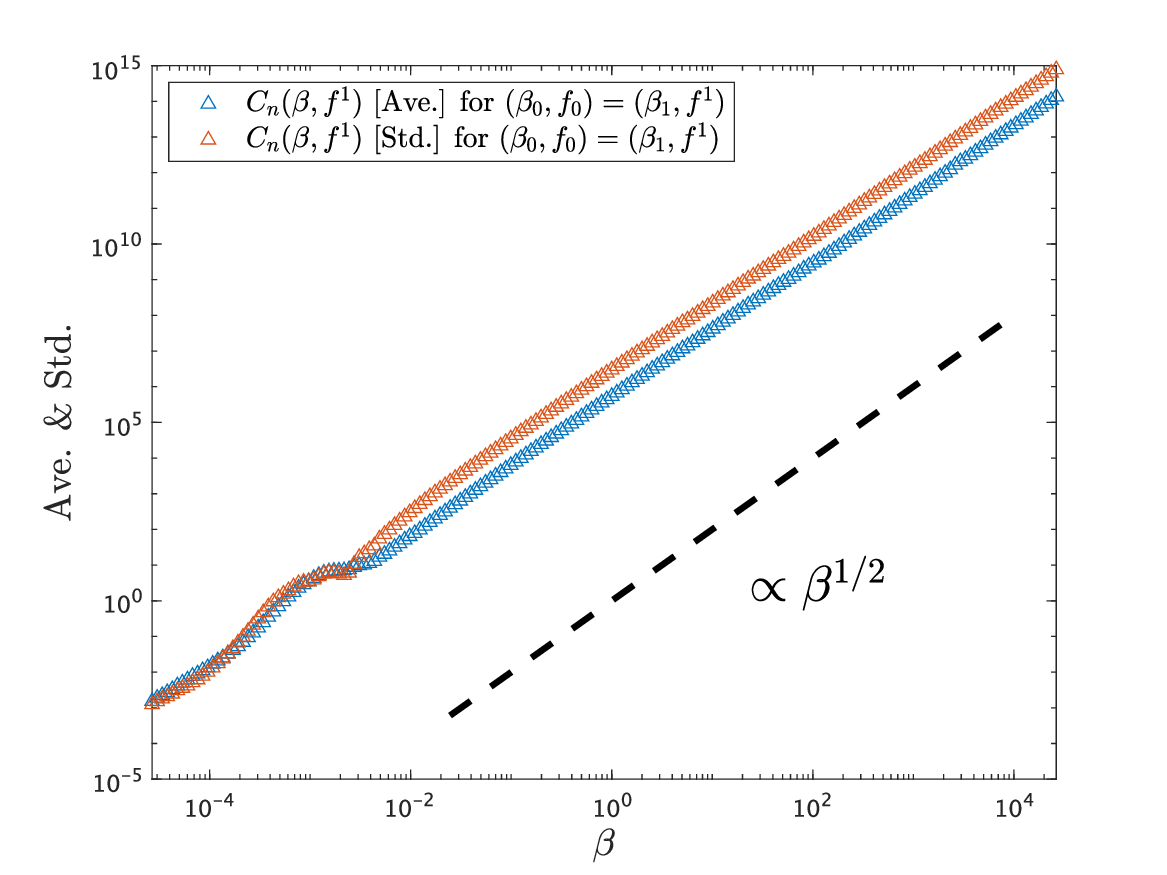}
&
\includegraphics[width=5.7cm]{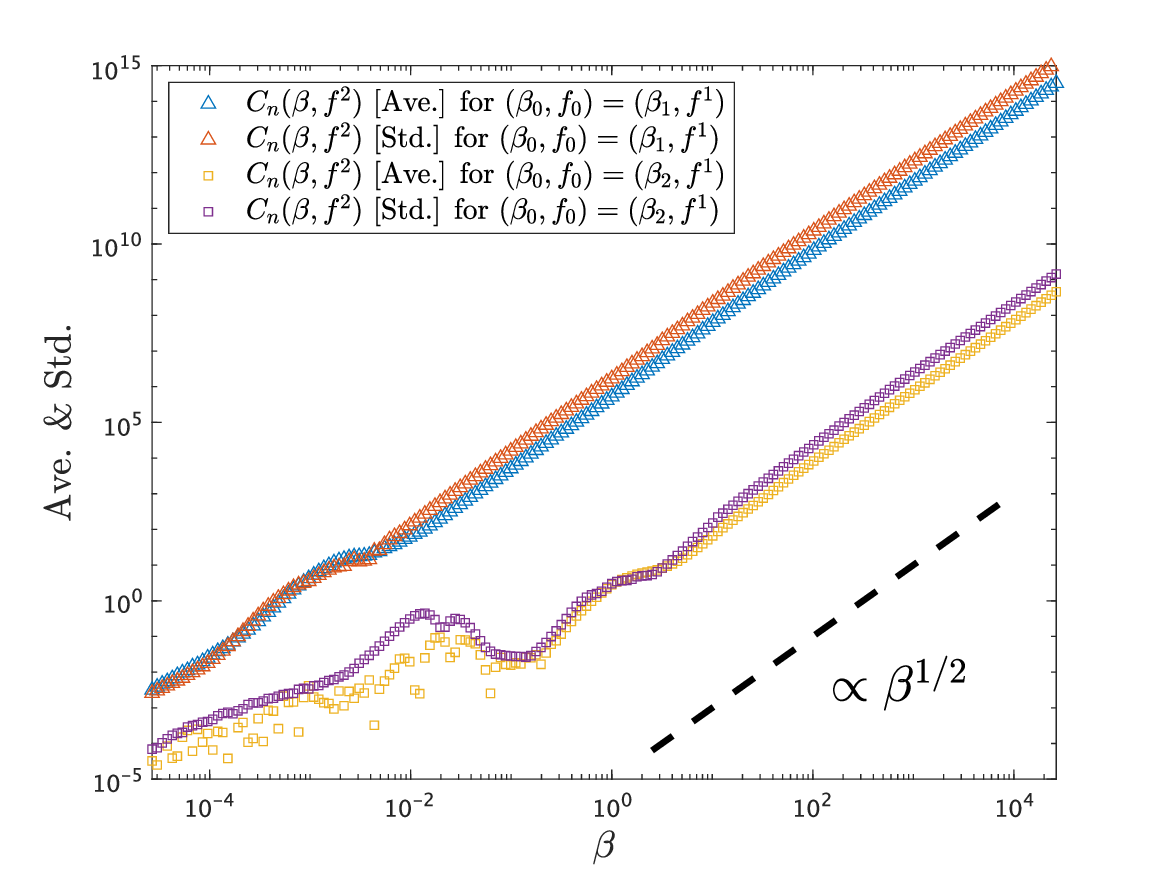} 
&
\includegraphics[width=5.7cm]{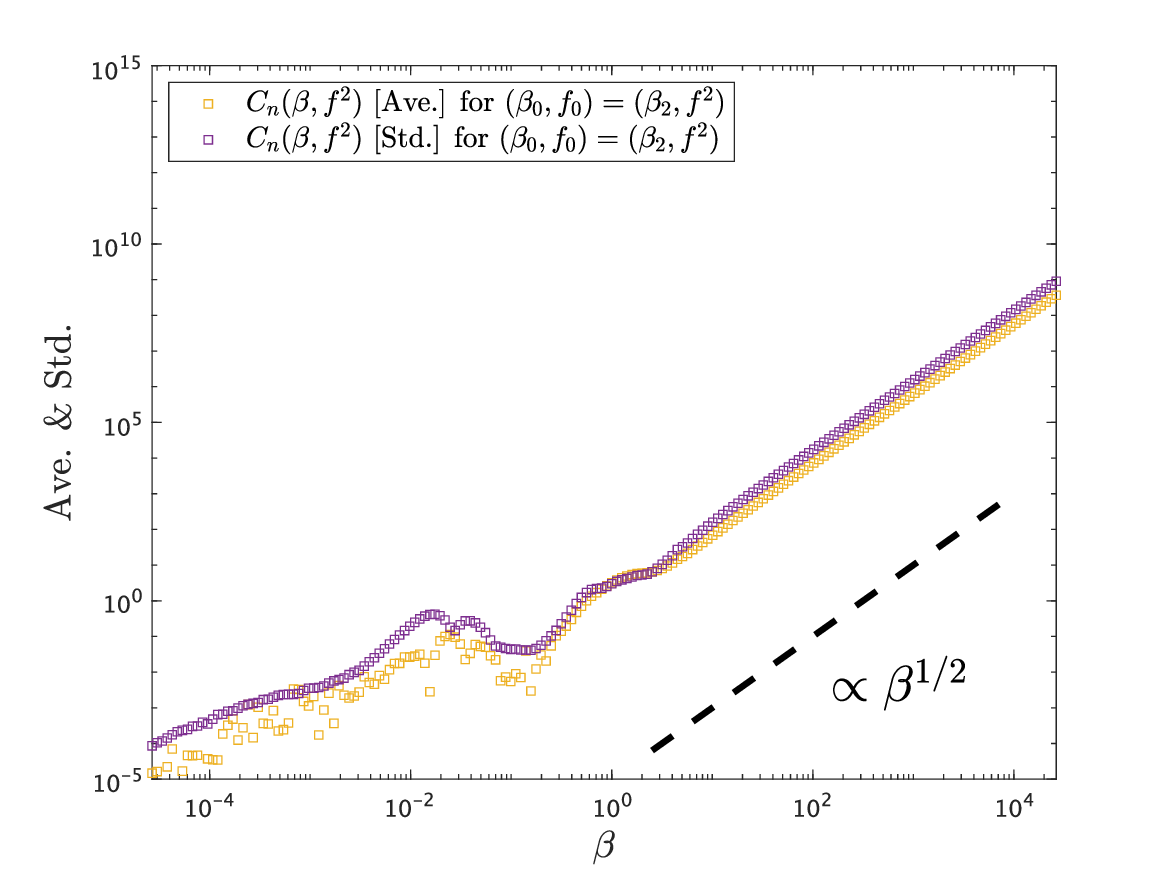}
\end{tabular}
\end{center}
\caption{Scaling analyses of the Bayes specific heat. Log-log plots of the expectations (blue triangles for $f_0=f^1$ and yellow squares for $f_0=f^2$) of $C_n(\beta,f)$, subtracting $\Lambda(\kappa, f; w_0, f_0)$ as the baseline, and the standard deviations (red triangles for $f_0=f^1$ and yellow squares for $f_0=f^2$) of $C_n(\beta,f)$ over realizations of $D^n$ in the cases of ({\bf a}) Fig. \ref{fig:Fig.S1}{\bf a}, ({\bf b}) Fig. \ref{fig:Fig.S1}{\bf b} and ({\bf c}) Fig. \ref{fig:Fig.S1}{\bf d}.}
\label{fig:Fig.S3}
\end{figure*}

\end{document}